\documentclass[12pt, draftcls, onecolumn]{IEEEtran}

\usepackage{xspace,amsfonts, amssymb, bbm, subfigure, graphicx,xcolor}
\usepackage[cmex10]{amsmath}
\usepackage{abbrevs}

\newcommand \E[1] {{\mathbbm{E}\left[{#1}\right]}}

\newcommand \Eh {\mathbbm{E}_\mathbf{h}}
\newcommand \Ez {\mathbbm{E}_\mathbf{z}}
\newcommand \Ezz[1] {{\Ez\left[{#1}\right]}}

\newcommand \boB {\mathbf{b}}
\newcommand \Pb[1] {#1}

\renewcommand \boB {B} 						%%% ESTAS DOS LINEAS HACEN QUE V SEA FUNCION DE B ESCALAR;
\renewcommand \Pb[1] {\big[#1\big]_2}		%%% SI SE COMENTAN V VUELVE A SER FUNCION DE b VECTOR

\newabbrev\CR{Cognitive radios (CRs)}[CR]
\newabbrev\SU{secondary users (SUs)}[SU]
\newabbrev\PU{primary users (PUs)}[PU]
\newabbrev\RA{\emph{resource allocation} (RA)}[RA]
\newabbrev\CSI{channel state information (CSI)}[CSI]
\newabbrev\DP{dynamic programming (DP)}[DP]
\newabbrev\POMDP{partially observable Markov decision process (POMDP)}[POMDP]
\newabbrev\NC{network controller (NC)}[NC]
\newabbrev\SIPN{state information of the primary network (SIPN)}[SIPN]
\newabbrev\SISN{state information of the secondary network (SISN)}[SISN]
\newabbrev\ROC{receiver operating characteristic (ROC)}[ROC]
\newabbrev\SNR{signal-to-noise ratio (SNR)}[SNR]
\newabbrev\QoS{quality of service (QoS)}[QoS]
\newabbrev\wrt{with respect to (w.r.t.)}[w.r.t.]
\newabbrev\IRI{\emph{instantaneous} reward indicators (IRIs)}[IRI]
\newabbrev\MDP{Markov Decision Processes (MDPs)}[MDP]

\makeatletter
\renewcommand\maybe@space@{%
  \maybe@ictrue % <= this is new
  \expandafter   \@tfor
    \expandafter \reserved@a
    \expandafter :%
    \expandafter =%
                 \nospacelist
                 \do \t@st@ic
  \ifmaybe@ic % <= this is new
    \space
  \fi
}
\makeatother

\begin{document}
\title{{\huge Jointly Optimal Sensing and Resource Allocation for Multiuser Overlay Cognitive Radios}}

\author{\normalsize{Luis M. Lopez-Ramos,
		Antonio G. Marques (contact author),
	and	Javier Ramos}		
\thanks{This work is supported by the Spanish Ministry of Science, under FPU Grant AP2010-1050. This paper has been
submitted for publication to the IEEE Journal on Selected Areas in
Communications. Parts of this paper were presented at CROWNCOM 2012.}
\thanks{All authors are with the Dept. of Signal Theory and Comms., King Juan Carlos Univ., Camino del Molino s/n, Fuenlabrada, Madrid 28943,
Spain. Contact info available at: http://www.tsc.urjc.es/}%
}

%\markboth{IEEE Journal on Selected Areas in Communications
%,~Vol.~XX, No.~XX, Month~2012}%
%(submitted)}%
%{Lopez-Ramos \MakeLowercase{\textit{et al.}}: Jointly Optimal Sensing and Resource Allocation for Multiuser Overlay Cognitive Radios}

\maketitle
\vspace{-1.5cm}
\begin{abstract} %\boldmath
Successful deployment of cognitive radios requires efficient sensing of the spectrum and dynamic adaptation of the available resources according to the sensed (imperfect) information. While most works design these two tasks separately, in this paper we address them jointly. In particular, we investigate an overlay cognitive radio with multiple secondary users that access orthogonally a set of frequency bands originally devoted to primary users. The schemes are designed to minimize the cost of sensing, maximize the performance of the secondary users (weighted sum rate), and limit the probability of interfering the primary users. The joint design is addressed using dynamic programming and nonlinear optimization techniques. A two-step strategy that first finds the optimal resource allocation for any sensing scheme and then uses that solution as input to solve for the optimal sensing policy is implemented. The two-step strategy is optimal, gives rise to intuitive optimal policies, and entails a computational complexity much lower than that required to solve the original formulation.
\end{abstract}
\vspace{-0.5cm}

\section{Introduction} \label{s:introduction}

{\CR} are viewed as the next-generation solution to alleviate the perceived spectrum scarcity. When {\CR}s are deployed, the {\SU} have to sense their radio environment to optimize their communication performance while avoiding (limiting) the interference to the {\PU}. As a result, effective operation of {\CR}s requires the implementation of two critical tasks: i) sensing the spectrum and ii) dynamic adaptation of the available resources according to the sensed information \cite{Haykin05}. To carry out the \emph{sensing task} two important challenges are: C1) the presence of errors in the measurements that lead to errors on the channel occupancy detection and thus render harmless SU transmissions impossible; and C2) the inability to sense the totality of the time-frequency lattice due to scarcity of resources (time, energy, or sensing devices). Two additional challenges that arise to carry out the \RA \emph{task} are: C3) the need of the \RA algorithms to deal with channel imperfections; and C4) the selection of metrics that properly quantify the reward for the {\SU}s and the damage for the {\PU}s.

Many alternatives have been proposed in the \CR literature to deal with these challenges. Different forms of imperfect
\CSI, such as quantized or noisy \CSI, have been used to deal with C1 \cite{refInterf2}. However, in the context of \CR, fewer works have considered the fact that the \CSI may be not only noisy but also outdated, or have incorporated those imperfections into the design of \RA algorithms \cite{zhao_separation}. The inherent trade-off between sensing cost and throughput gains in C2 has been investigated \cite{liang_ST_tradeoff}, and designs that account for it based on convex optimization \cite{xin_convex} and \DP \cite{zhao_separation} for specific system setups have been proposed. Regarding C3, many works consider that the \CSI is imperfect, but only a few exploit the statistical model of these imperfections (especially for the time correlation) to mitigate them; see, e.g., \cite{zhao_separation,JSAC}.
Finally, different alternatives have been considered to deal with C4 and limit the harm that the {\SU}s cause to the {\PU}s \cite{Sergiy_icassp11}. The most widely used is to set limits on the peak (instantaneous) and average interfering power. Some works also have imposed limits on the rate loss that {\PU}s experience \cite{quantizedCR,marques_cap_crown12}, while others look at limiting the instantaneous or average probability of interfering the {\PU} (bounds on the short-term or long-term outage probability)  \cite{NeelyCR09,JSAC}.

Regardless of the challenges addressed and the formulation chosen, the sensing and \RA policies have been traditionally designed separately. Each of the tasks has been investigated thoroughly and relevant results are available in the literature. However, a globally optimum design requires designing those tasks jointly, so that the interactions among them can be properly exploited. Clearly, more accurate sensing enables more efficient \RA, but at the expense of higher time and/or energy consumption. Early works dealing with joint design of sensing and \RA are \cite{zhao_framework} and \cite{zhao_separation}. In such works, imperfections in the sensors, and also time correlation of the state of the primary channel, are considered. As a result, the sensing design is modeled as a \POMDP \cite{pomdpsurvey}, which can be viewed as a specific class of \DP. The design of the \RA in these works amounts to select the user  transmitting on each channel (also known as user scheduling). Under mild conditions, the authors establish that a separation principle holds in the design of the optimal access and sensing policies. Additional works addressing the joint design of sensing and \RA, and considering more complex operating conditions, have been published recently \cite{xin_convex,kim_sequential}. For a single {\SU} operating multiple fading channels, \cite{xin_convex} relies on convex optimization to optimally design both the \RA and the indexes of the channels to be sensed at every time instant. Assuming that the number of channels that can be sensed at every instant is fixed and that the primary activity is independent across time, the author establishes that the channels to sense are the ones that can potentially yield a higher reward for the secondary user. Joint optimal design is also pursued in \cite{kim_sequential}, although for a very different setup. Specifically, \cite{kim_sequential} postulates that at each slot, the \CR must calculate the fraction of time devoted to sense the channel and the fraction devoted to transmit in the bands which are found to be unoccupied. Clearly, a trade-off between sensing accuracy and transmission rate emerges. The design is formulated as an optimal stopping problem, and solved by means of Lagrange relaxation of \DP \cite{castanon}. However, none of these two works takes into account the temporal correlation of the \SIPN.

The objective of this work is to design the sensing and the \RA policies \emph{jointly} while accounting for the challenges C1-C4.
The specific operating conditions considered in the paper are described next. We analyze an \emph{overlay}\footnote{Some authors refer to overlay networks as interweave networks, see, e.g., \cite{GoldsmithCR}.} \CR with multiple {\SU}s and {\PU}s. {\SU}s are able to adapt their power and rate loadings and access orthogonally a set of frequency bands. Those bands are originally devoted to {\PU}s transmissions. \emph{Orthogonally} here means that if a {\SU} is transmitting, no other {\SU} can be active in the same band. The schemes are designed to maximize the sum-average rate of the {\SU}s while adhering to constraints that \emph{limit} the maximum ``average power'' that {\SU}s transmit and the average ``probability of interfering'' the {\PU}s. It is assumed that the \CSI of the {\SU} links is instantaneous and free of errors, while the \CSI of the {\PU}s activity is outdated and noisy. A simple first-order hidden Markov model is used to characterize such imperfections. Sensing a channel band entails a given cost, and at each instant the system has to decide which channels (if any) are sensed.

The jointly optimal sensing and \RA schemes will be designed using \DP and nonlinear optimization techniques. \DP techniques are required because the activity of {\PU}s is assumed to be correlated across time, so that sensing a channel has an impact not only for the current instant, but also for future time instants \cite{zhao_framework}. To solve the joint design, a two-step strategy is implemented. In the first step, the sensing is considered given and the optimal \RA is found for \emph{any} fixed sensing scheme. This problem was recently solved in \cite{marques_icassp, JSAC}. In the second step, the results of the first step are used as input to obtain the optimal sensing policy. The motivation for using this two-step strategy is twofold. First, while the joint design is non convex and has to be solved using \DP techniques, the problem in the first step (optimal \RA for a fixed sensing scheme) can be recast as a convex one. Second, when the optimal \RA is substituted back into the original joint design, the resulting problem (which does need to be solved using \DP techniques) has a more favorable structure. More specifically, while the original design problem was a constrained \DP, the updated one is an unconstrained \DP problem which can be solved separately for each of the channels.

The rest of the paper is organized as follows. Sec. II describes the system setup and introduces notation. The optimization problem that gives rise to the optimal sensing and \RA schemes is formulated in Sec. \ref{s:problem_statement}. The solution for the optimal \RA given the sensing scheme is presented in Sec. \ref{s:RA}. The optimization of the sensing scheme is addressed in Sec. \ref{s:sensing}. The section begins with a brief review of \DP and {\POMDP}s. Then, the problem is formulated in the context of \DP and its solution is developed. Numerical simulations validating the theoretical claims and providing insights on our optimal schemes are presented in Sec. \ref{s:simulations}. Sec. \ref{s:analyzing_future} analyzes the main properties of our jointly optimal \RA and sensing policies, provides insights on the operation of such policies, and points out future lines of work.\footnote{ \emph{Notation:} $x^*$ denotes the optimal value of variable $x$; $\mathbbm{E}[\cdot]$ expectation; $\wedge$ the boolean ``and'' operator; $\mathbbm{1}_{\{\cdot\}}$
the indicator function ($\mathbbm{1}_{\{x\}}=1$ if $x$ is true and
zero otherwise); and $[x]_+$ the projection of $x$ onto the non-negative orthant, i.e., $[x]_+:=\max\{x,0\}$.}

\section{System setup and state information} \label{s:SystemSetup}

This section is devoted to describe the basic setup of the system. We begin by briefly describing the system setup and the operation of the system (tasks that the system runs at every time slot). Then, we explain in detail the model for the \CSI, which will play a critical role in the problem formulation. The resources that {\SU}s will adapt as a function of the \CSI are described in the last part of the section.

We consider a \CR scenario with several {\PU}s and {\SU}s. The frequency band of interest (the portion of spectrum that is licensed to {\PU}s, or the subset of this shared with the {\SU}s, if not all) is divided into $K$ frequency-flat orthogonal subchannels (indexed by $k$). Each of the $M$ secondary users (indexed by $m$) opportunistically accesses any number of these channels during a time slot (indexed by $n$). Opportunistic here means that the user accessing each channel will vary with time, with the objective of optimally utilizing the available channel resources. For simplicity, we assume that there is a \NC which acts as a central scheduler and will also perform the task of sensing the medium for primary presence. The scheduling information will be forwarded to the mobile stations through a parallel feedback channel. The results hold for one-hop (either cellular or any-to-any) setups.

Next, we briefly describe the operation of the system. A more detailed description will be given in Sec. \ref{s:problem_statement}, which will rely on the notation and problem formulation introduced in the following sections. Before starting, it is important to clarify that we focus on systems where the \SIPN is more difficult to acquire than the \SISN. As a result, we will assume that \SISN is error-free and acquired at every slot $n$, while \SIPN is not. With these considerations in mind, the \CR operates as follows. At every slot $n$ the following tasks are run sequentially: T1) the \NC acquires the \SISN; T2) the \NC relies on the output of T1 (and on previous measurements) to decide which channels to sense (if any), then the output of the sensing is used to update the \SIPN; and T3) the \NC uses the outputs of T1 and T2 to find the optimal \RA for instant $n$. Overheads associated with acquisition of the \SISN and notification of the optimal \RA to the {\SU}s are considered negligible. Such an assumption facilitates the analysis, and it is reasonable for scenarios where the {\SU}s are deployed in a relatively small area which allows for low-cost signaling transmissions.

\subsection{State information and sensing scheme}\label{ss:SystemSetup:CSI}

We begin by introducing the model for the \SISN. The noise-normalized square magnitude of the fading coefficient (power gain) of the channel between the $m$th secondary user and its intended receiver on frequency $k$ during slot $n$ is denoted as $h_k^m[n]$. Channels are random, so that $h_k^m[n]$ is a stochastic process, which is assumed to be independent across time. The values of $h_m^k[n]$ for all $m$ and $k$ form the \SISN at slot $n$. We assume that the \SISN is perfect, so that the values of $h_m^k[n]$ at every time slot $n$ are know perfectly (error-free). While \SISN comprises the power gains of the secondary links, the \SIPN accounts for the channel occupancy. We will assume that the primary system contains one user per channel. This assumption is made to simplify the analysis and it is reasonable for certain primary systems, e.g. mobile telephony where a single narrow-band channel is assigned to a single user during the course of a call. Since we consider an overlay scenario, it suffices to know whether a given channel is occupied or not \cite{GoldsmithCR}. This way, when a {\PU} is not active, opportunities for {\SU}s to transmit in the corresponding channel arise. The primary system is not assumed to collaborate with the secondary system. Hence, from the point of view of the {\SU}s, the behavior of {\PU}s is a stochastic process independent of $h_k^m[n]$. With these considerations in mind, the presence of the primary user in channel $k$ at time $n$ is represented by the binary state variable $a_k[n]$ (0/idle, 1/busy). Each primary user's behavior will be modeled as a simple Gilbert-Elliot channel model, so that $a_k[n]$ is assumed to remain constant during the whole time slot, and then change according to a two-state, time invariant Markov chain. The Markovian property will be useful to keep the \DP modeling simple and will also be exploited to recursively keep track of the \SIPN. Nonetheless, more advanced models can be considered without paying a big computational price \cite{JSAC,nonMarkovPU_Act}. With $P_k^{xy} := \Pr (a_k[n]=x\vert a_k[n-1]=y)$, the dynamics for the Gilbert-Elliot model are fully described by the $2\times 2$ Markov transition matrix $\mathbf{P}_k := [P_k^{00}, P_k^{01}; P_k^{10}, P_k^{11}]$. Sec. \ref{s:analyzing_future} discusses the implications of relaxing some of these assumptions.

%It is realistic to consider that it is impossible to have an error-free sensing. Such a task is affected by the SNR of the channel between primary and secondary users, and the knowledge of the statistical properties of the primary signals. There are three classic approaches to signal detection: matched filter, energy detection and cyclostationary feature detection. These three techniques exhibit a trade-off between noise sensitivity and required prior knowledge of the channel usage.

%Similarly to [zhao-framework], a Neymann-Pearson criterion is used to choose this operation point. The $P^{MD}_k$ is constrained to be less than or equal to the maximum allowed probability of interference $\check{o}_k$ and the $P^{FA}_k$ is minimized subject to this constraint. Due to the aforementioned sources of imperfections,

While knowledge of $h_k^m[n]$ at instant $n$ was assumed to be perfect (deterministic), knowledge of $a_k[n]$ at instant $n$ is assumed to be imperfect (probabilistic). Two important sources of imperfections are: i) errors in the sensing process and ii) outdated information (because the channels are not always sensed). For that purpose, let $s_k[n]$ denote a binary design variable which is 1 if the $k$th channel is sensed at time $n$, and 0 otherwise. Moreover, let $z_k[n]$ denote the output of the sensor if indeed $s_k[n]=1$; i.e., if the $k$th channel has been sensed. We will assume that the output of the sensor is binary and may contain errors. To account for asymmetric errors, the probabilities of false alarm $P^{FA}_k = \Pr(z_k[n]=1\vert a_k[n]=0)$ and miss detection $P^{MD}_k = \Pr(z_k[n]=0\vert a_k[n]=1)$ are considered.  Clearly, the specific values of $P^{FA}_k$ and $P^{MD}_k$ will depend on the detection technique the sensors implement (matched filter, energy detector, cyclostationary detector, etc.) and the working point of the \ROC curve, which is usually controlled by selecting a threshold \cite{arslan_sensing}. In our model, this operation point is chosen beforehand and it is fixed during the system operation, so that the values of $P^{FA}_k$ and $P^{MD}_k$ are assumed known. As already mentioned, the sensing imperfections render the knowledge of $a_k[n]$ at instant $n$ probabilistic. In other words, $a_k[n]$ is a partially observable state variable. The knowledge about the value of $a_k[n]$ at instant $n$ will be referred to as (instantaneous) belief,
also known as the information process. For a given instant $n$, two different beliefs are considered: the \emph{pre-decision} belief $B_k[n]$ and the \emph{post-decision} belief $B_k^S[n]$. Intuitively, $B_k[n]$ contains the information about $a_k[n]$ before the sensing decision has been made (i.e., at the beginning of task T2), while $B_k^S[n]$ contains the information about $a_k[n]$ once $s_k[n]$ and $z_k[n]$ (if $s_k[n]=1$) are known (i.e., at the end of task T2). Mathematically, if $\mathcal{H}_n$ represents the history of all sensing decisions and measurements, i.e., $\mathcal{H}_n:=\{s_k[0],z_k[0],\ldots,s_k[n],z_k[n]\}$; then $B_k[n] := \Pr(a_k[n]=1 \vert \mathcal{H}_{n-1} )$ and $B_k^S[n] := \Pr(a_k[n]=1 \vert \mathcal{H}_{n})$. For notational convenience, the beliefs will also be expressed as vectors, with $\mathbf{b}_k[n] := \Big[ 1-B_k[n], B_k[n]\Big]^T$ and $\mathbf{b}^S_k[n] = \Big[1- B^S_k[n], B^S_k[n]\Big]^T$. Using basic results from Markov chain theory and provided that $\mathbf{P}_k$ (time-correlation model) is known, the expression to get the pre-decision belief at time slot $n$ is
\begin{equation}\label{E:gilbert_predict}
\mathbf{b}_k[n]= \mathbf{P}_k\mathbf{b}_k^S[n-1].
\end{equation}
Differently, the expression to get $\mathbf{b}_k^S[n]$ depends on the sensing decision $s_k[n]$. If $s_k[n]=0$, no additional information is available, so that
\begin{equation} \label{E:gilbert_correct_s0}
\mathbf{b}_k^S[n] = \mathbf{b}_k[n].
\end{equation}
If $s_k[n]=1$, the belief is corrected as $\mathbf{b}_k^S[n]=\mathbf{b}_k^S \Big(\mathbf{b}_k[n], z_k[n]\Big)$, with
\begin{equation} \label{E:gilbert_correct_s1}
\mathbf{b}_k^S \Big(\mathbf{b}_k[n], z\Big) :=
\frac{\mathbf{D}_{z}\mathbf{b}_k[n]}{\Pr(z_k[n] = z \big\vert \mathbf{b}_k[n])}, \hspace*{-0.5mm}
\end{equation}
where $\mathbf{D}_{z}$ with $z\in\{0,1\}$ is a $2\times2$ diagonal matrix with entries $[\mathbf{D}_{z}]_{1,1}:=\Pr(z_k[n]=z\vert a_k=0)$ and $[\mathbf{D}_{z}]_{2,2}:=\Pr(z_k[n]=z \vert a_k=1)$. Note that the denominator is the probability of an outcome conditioned to a specified belief: $ \Pr(z_k[n]=z \big\vert \mathbf{b}_k[n]) = \mathbf{1}^T\mathbf{D}_{z}\mathbf{b}_k[n] $, so that \eqref{E:gilbert_correct_s1} corresponds to the correction step of a Bayesian recursive estimator. If no information about the initial state of the {\PU} is available, the best choice is to initialize $\mathbf{b}_k[0]$ to the stationary distribution of the Markov chain associated with channel $k$ (i.e., the principal eigenvector of $\mathbf{P}_k$).

In a nutshell, the actual state of the primary and secondary networks is given by the random processes $a_k[n]$ and $h_k^m[n]$, which are assumed to be independent. The operating conditions of our \CR are such that at instant $n$, the value of $h_k^m[n]$ is perfectly known, while the value of $a_k[n]$ is not. As a result, the \SIPN is not formed by $a_k[n]$, but by $\mathbf{b}_k[n]$ and $\mathbf{b}_k^S[n]$ which are a probabilistic description of $a_k[n]$. The system will perform the sensing and \RA tasks based on the available \SISN and \SIPN. In particular, the sensing decision will be made based on $h_k^m[n]$ and $\mathbf{b}_k[n]$, while the \RA will be implemented based on $h_k^m[n]$ and $\mathbf{b}_k^S[n]$.

%vectors $\mathbf{b}_k[n]$ and $\mathbf{b}_k^S[n]$ (for all $k$) and channel matrix $\mathbf{H}[n]$ are gathered into the vector $\mathbf{i}[n]$, which represents the overall \CSI of the system.

%Proceso de sensado. Belief
%The decision of sensing or not a channel at slot $n$ is denoted as $s_k[n]$. The sensing decision is made at the beginning of the time slot, in such a way that the outcome of the sensing can be used to improve the scheduling in the same slot. In order to keep the information gathered by past measures on channel $k$, we use a variable called belief $\mathbf{b}_k[n]$. This variable is a vector that contains the posterior probabilities that the channel is idle and busy, conditioned to all past measures: $\mathbf{b}_k[n] = [ Pr(a_k[n]=0 \vert z_k[0], z_k[1], ... z_k[n-1] ), Pr(a_k[n]=1 \vert z_k[0], z_k[1], ... z_k[n-1] )]$. The belief can be updated recursively, so that there is no need to keep in memory all past measures, it is enough with $\mathbf{b}_k[n] $.

\subsection{Resources at the secondary network}\label{ss:SystemSetup:ResourcesSN}

We consider a secondary network where users are able to implement adaptive modulation and power control, and share orthogonally the available channels. To describe the channel access scheme (scheduling) rigorously, let $w_k^m[n]$ be a boolean variable so that $w_k^m[n]=1$ if {\SU} $m$ accesses channel $k$ and zero otherwise. Moreover, let $p_k^m[n]$ be a nonnegative variable denoting the nominal power {\SU} $m$ transmits in channel $k$, and let $C_k^m[n]$ be its corresponding rate. We say that the $p_m^k[n]$ is a nominal power in the sense that power is consumed only if the user is actually accessing the channel. Otherwise the power is zero, so that the actual (effective) power user $m$ loads in channel $k$ can be written as $w_k^m[n]p_k^m[n]$.

The transmission bit rate is obtained through Shannon's capacity formula \cite{Li-Goldsmith01a}: $C_k^m[n]:=\log_2(1+h_k^m[n] p_k^m[n]/\Gamma)$ where $\Gamma$ is a \SNR gap that accounts for the difference between the theoretical capacity and the actual rate achieved by the modulation and coding scheme the {\SU} implements. This is a bijective, nondecreasing, concave function with $p_k^m[n]$ and it establishes a relationship between power and rate in the sense that controlling $p_k^m[n]$ implies also controlling $C_k^m[n]$.

The fact of the access being orthogonal implies that, at any time instant, at most one {\SU} can access the channel. Mathematically,
\begin{equation}\label{E:c_inst_sched}
\sum_m w_k^m[n]\leq 1 \; \; \forall k,n.
\end{equation}
Note that \eqref{E:c_inst_sched} allows for the event of all $w_k^m[n]$ being zero for a given channel $k$. That would happen, if, for example, the system thinks that it is very likely that channel $k$ is occupied by a {\PU}.

\section{Problem statement} \label{s:problem_statement} % Secci—n III

%The following section is a summary of the design presented in [marques-ICASSP] and stresses the most important results for the problem to solve in this work.
The approach in this paper is to design the sensing and \RA schemes as the solution of a judiciously formulated optimization problem. Consequently, it is critical to identify: i) the design (optimization) variables, ii) the state variables, iii) the constraints that design and state variables must obey, and iv) the objective of the optimization problem.

The first two steps were accomplished in the previous section, stating that the design variables are $s_k[n]$, $w_k^m[n]$ and $p_k^m[n]$ (recall that there is no need to optimize over $C_k^m[n]$); and that the state variables are $h_k^m[n]$ (\SISN), and  $\mathbf{b}_k[n]$ and $\mathbf{b}_k^S[n]$ (\SIPN).

Moving to step iii), the constraints that the variables need to satisfy can be grouped into two classes. The first class is formed by constraints that account for the system setup. This class includes constraint \eqref{E:c_inst_sched} as well as the following constraints that were implicitly introduced in the previous section: $s_k[n]\in\{0,1\}$, $w_k^m[n]\in\{0,1\}$ and $p_k^m[n]\geq 0$. The second class is formed by constraints that account for \QoS. In particular, we consider the following two constraints. The first one is a limit on the maximum average (long-term) power a \SU can transmit. By enforcing an average consumption constraint, opportunistic strategies are favored because energy can be saved during deep fadings (or when the channel is known to be occupied) and used during transmission opportunities. Transmission opportunities are time slots where the channel is certainly known to be idle and the fading conditions are favorable. Mathematically, with $\check{p}^m$ denoting such maximum value, the average power constraint is written as:
\begin{equation}\label{E:c_power}
\mathbb{E}\left[  \lim_{N \to \infty} (1-\gamma)\sum_{n=0}^{N-1} \gamma^n \sum\nolimits_{ k}w_k^m[n]p_k^m[n]\right]\leq\check{p}^m,~~\forall m,
\end{equation}
where $0 < \gamma < 1$ is a discount factor such that more emphasis is placed in near future instants. The factor $(1-\gamma)$ ensures that the averaging operator is normalized; i.e., that $\lim_{N \to \infty}\sum_{n=0}^{N-1} (1-\gamma)\gamma^n=1$. As explained in more detail in Sec. \ref{s:sensing}, using an exponentially decaying average is also useful from a mathematical perspective (convergence and existence of stationary policies are guaranteed).

While the previous constraint guarantees \QoS for the {\SU}s, we also need to guarantee a level of \QoS for the {\PU}s. As explained in the introduction, there are different strategies to limit the interference that {\SU}s cause to {\PU}s; e.g., by imposing limits on the interfering power at the {\PU}s, or on the rate loss that such interference generates \cite{JSAC}. In this paper, we will guarantee that the \emph{long-term} probability of a {\PU} being interfered by {\SU}s is below a certain prespecified threshold $\check{o}_k$. Mathematically, we require $\Pr\{\sum\nolimits_{m}w_k^m =1|a_k =1\}\leq \check{o}_k$ for each band $k=1,\ldots,K$. Using Bayes' theorem, and capitalizing on the fact that both $a_k$ and $\sum\nolimits_{m}w_k^m$ are boolean variables, the constraint can be re-written as:
\begin{equation}\label{E:c_prob_int}
\mathbbm{E}\left[  \lim_{N \to \infty} \sum_{n=0}^{N-1} (1-\gamma)\gamma^n a_k[n] \sum\nolimits_{m}w_k^m[n]\right]/A_k\leq\check{o}_k,~~\forall k,
\end{equation}
where $A_k$, which is assumed known, denotes the stationary probability of the $k$th band being occupied by the corresponding primary user. Writing the constraint in this form reveals its underlying convexity.
Before moving to the next step, two clarifications are in order. The first one is on the practicality of \eqref{E:c_prob_int}. Constraints that allow for a certain level of interference are reasonable because error-free sensing is unrealistic. Indeed, our model assumes that even if channel $k$ is sensed as idle, there is a probability $P_k^{MD}$ of being occupied. Moreover, when the interference limit is formulated as long-term constraint (as it is in our case), there is an additional motivation for the constraint. The system is able to exploit the so-called interference diversity \cite{interference_diversity}. Such diversity allows {\SU}s to take advantage of very good channel realizations even if they are likely to interfere {\PU}s. To balance the outcome, {\SU}s will be conservative when channel realizations are not that good and may remain silent even if it is likely that the {\PU} is not present. The second clarification is that we implicitly assumed that {\SU} transmissions are possible even if the {\PU} is present. The reason is twofold. First, the fact that a {\SU} transmitter is interfering a {\PU} receiver, does not necessarily imply that the reciprocal is true. Second, since the \NC does not have any control over the power that primary transmitters use, the interfering power at the secondary receiver is a state variable. As such, it could be incorporated into $h_k^m[n]$ as an additional source of noise.

The fourth (and last) step to formulate the optimization problem is to design the metric (objective) to be maximized. Different utility (reward) and cost functions can be used to such purpose. As mentioned in the introduction, in this work we are interested in schemes that maximize the weighted sum rate of the {\SU}s and minimize the cost associated with sensing. Specifically, we consider that every time that channel $k$ is sensed, the system has to pay a price $\xi_k>0$. We assume that such a price is fixed and known beforehand, but time-varying prices can be accommodated into our formulation too (see Sec. \ref{ss:analyzing_future:sensing_cost} for additional details). This way, the sensing cost at time $n$ is $U_S[n]:=\sum_{k}\xi_k s_k[n]$. Similarly, we define the utility for the {\SU}s at time $n$ as $U_{SU}[n] :=\sum_{k} \left(\sum_{m} \beta^m w_k^m[n] C_k^m(h_k^m[n],p_k^{m}[n]) \right)$, where $\beta^m>0$ is a user-priority coefficient. Based on these definitions, the utility for our \CR at time $n$ is $U_T[n]:= U_{SU}[n]-U_S[n]$. Finally, we aim to maximize the long-term utility of the system denoted by $\bar{U}_T$ and defined as
\begin{equation} \label{E:ubar}
\bar{U}_T := \mathbbm{E} \left[ \lim_{N \to \infty} \sum_{n=0}^{N-1} (1-\gamma)\gamma^n U[n] \right].
\end{equation}

With these notational conventions, the optimal $s_k^*[n]$, $w_k^{m*}[n]$ and $p_k^{m*}[n]$ will be obtained as the solution of the following constrained optimization problem.
\begin{subequations}\label{E:constrained_original}
	\begin{alignat}{2} \label{E:constrained_original_obj}
	\max_{\{s_k[n], w_k^m[n], p_k^m[n]\}} & ~~\bar{U}_T\\
       ~\mathrm{s.~to:}~~~~& ~\eqref{E:c_inst_sched},~ w_k^m[n]\in \{0,1\},~p_k^m[n]\geq 0 \label{E:constrained_original_shortterm_RA}\\
        & ~\eqref{E:c_power},~\eqref{E:c_prob_int} \label{E:constrained_original_longterm_RA}\\
       & ~ s_k[n]\in \{0,1\} \label{E:constrained_original_shortterm_Sensing}.
\end{alignat}
\end{subequations}
Note that constraints in \eqref{E:constrained_original_shortterm_RA} and \eqref{E:constrained_original_longterm_RA} affect the design variables involved in the \RA task ($w_k^m[n]$ and $p_k^m[n]$), while \eqref{E:constrained_original_shortterm_Sensing} affects the design variables involved in the sensing task ($s_k[n]$). Moreover, the reason for writing \eqref{E:constrained_original_shortterm_RA} and \eqref{E:constrained_original_longterm_RA} separately is that \eqref{E:constrained_original_shortterm_RA} refers to constraints that need to hold for each and every time instant $n$, while \eqref{E:constrained_original_longterm_RA} refers to constraints that need to hold in the long-term.

The main difficulty in solving \eqref{E:constrained_original} is that the solution for all time instants has to be found jointly. The reason is that sensing decisions at instant $n$ have an impact not only at that instant, but at future instants too. As a result, a separate per-slot optimization approach is not optimal, and \DP techniques have to be used instead. Since \DP problems generally have exponential complexity, we will use a two step-strategy to solve \eqref{E:constrained_original} which will considerably reduce the computational burden without sacrificing optimality. To explain such a strategy, it is convenient to further clarify the operation of the system. In Sec. \ref{s:SystemSetup} we explained that at each slot $n$, our \CR had to implement three main tasks: T1) acquisition of the \SISN, T2) sensing and update of the \SIPN, and T3) allocation of resources. In what follows, task T2 is split into 3 subtasks, so that the \CR runs five sequential steps:
\begin{itemize}
\item T1) At the beginning of the slot, the system acquires the exact value of the channel gains $h_k^m[n]$;
\item T2.1) the Markov transition matrix and the post-decision beliefs $\mathbf{b}_k^S[n-1]$ of the previous instant are used to obtain pre-decision beliefs $\mathbf{b}_k[n]$ via \eqref{E:gilbert_predict};
\item T2.2) $h_k^m[n]$ and $\mathbf{b}_k[n]$ are used to find $s_k^*[n]$;
\item T2.3) $s_k^*[n]$ and $z_k[n]$ (for the channels for which $s_k^*[n] = 1$) are used to get the post-decision beliefs $\mathbf{b}_k^S[n]$ via \eqref{E:gilbert_correct_s0} and \eqref{E:gilbert_correct_s1};
\item T3) $h_k^m[n]$ and $\mathbf{b}_k^S[n]$ are used to find the optimal value of $w_k^{m*}[n]$ and $p_k^{m*}[n]$, and the {\SU}s transmit accordingly.
\end{itemize}

The two-step strategy to solve \eqref{E:constrained_original} will proceed as follows. In the first step, we will find the optimal $w_k^m[n]$ and $p_k^m[n]$ for any sensing scheme. Such a problem is simpler than the original one in \eqref{E:constrained_original} not only because the dimensionality of the optimization space is smaller, but also because we can ignore (drop) all the terms in \eqref{E:constrained_original} that depend only on $s_k[n]$. This will be critical, because if the sensing is not optimized, a per-slot optimization \wrt the remaining design variables is feasible. In the second step, we will substitute the output of the first step into \eqref{E:constrained_original} and solve for the optimal $s_k[n]$. Clearly, the output of the first step will be used in T3 while the output of the second step will be used in T2.2. The optimization in the first step (\RA) is addressed next, while the optimization in the second step (sensing) is addressed in Sec. \ref{s:sensing}.

\section{Optimal RA for the secondary network}\label{s:RA} % Secci—n IV

According to what we just explained, the objective of this section is to design the optimal \RA (scheduling and powers) for a fixed sensing policy. It is worth stressing that solving this problem is convenient because: i) it corresponds to one of the tasks our \CR has to implement; ii) it is a much simpler problem than the original problem in \eqref{E:constrained_original}, indeed the problem in this section has a smaller dimensionality and, more importantly, can be recast as a convex optimization problem; and iii) it will serve as an input for the design of the optimal sensing, simplifying the task of finding the global solution of \eqref{E:constrained_original}.

Because in this section the sensing policy is considered given (fixed), $s_k[n]$ is not a design variable, and all the terms that depend only on $s_k[n]$ can be ignored. Specifically, the sensing cost $U_S[n]$ in \eqref{E:constrained_original_obj} and the constraint in \eqref{E:constrained_original_shortterm_Sensing} can be dropped. The former implies that the new objective to optimize is $\bar{U}_{SU}:=\sum\nolimits_{ k,m}\mathbbm{E}[\lim_{N \to \infty} \sum_{n=0}^{N-1} (1-\gamma)
       \gamma^n \beta^mw_k^m[n] C_k^m(h_k^m[n],p_k^{m}[n])]$. With these considerations in mind, we aim to solve the following problem [cf. \eqref{E:constrained_original}]
\begin{subequations}\label{E:optimization_problem}
\begin{alignat}{2} \label{E:optRA_obj}
        \hspace{-.3cm}\underset{\{w_k^m[n],p_k^m[n]\}}{\max}
       ~~&\bar{U}_{SU}\hspace{5.04cm}\\
               \label{E:optRA_const}~\mathrm{s.~to:}
               ~~&\eqref{E:constrained_original_shortterm_RA},~ \eqref{E:constrained_original_longterm_RA}.
\end{alignat}
\end{subequations}

A slightly modified version of this problem was recently posed and solved in \cite{JSAC}. For this reason we organize the remaining of this section into two parts. The first one summarizes (and adapts) the results in \cite{JSAC}, presenting the optimal \RA. The second part is devoted to introduce new variables that will serve as input for the design of the optimal sensing in Sec. \ref{s:sensing}.

\subsection{Solving for the RA}\label{ss:RA:presenting_opt_RA}

It can be shown that after introducing some auxiliary (dummy) variables and relaxing the constraint $w_k^m[n] \in \{0,1\}$ to $w_k^m[n] \in [0,1]$, the resultant problem in \eqref{E:optimization_problem} is convex. Moreover, with probability one the solution to the relaxed problem is the same than that of the original problem; see \cite{JSAC} as well as \cite{amggjr_tsp11} for details on how to obtain the solution for this problem. The approach to solve \eqref{E:optimization_problem} is to dualize the long-term constraints in \eqref{E:constrained_original_longterm_RA}. For such a purpose, let $\pi^m$ and $\theta_k$ be the Lagrange multipliers associated with constraints \eqref{E:c_power} and \eqref{E:c_prob_int}, respectively. It can be shown then that the optimal solution to \eqref{E:optimization_problem} is
\begin{eqnarray}
\label{E:opt_pow}p_k^{m*}[n] &:=& %p = C^(-1)(h, pi)
	\left[(\dot{C}_k^m)^{-1}\left(h_k^m[n],\pi^m/\beta^m\right)\right]_+;\hspace{.3cm}\\
\label{E:opt_sched}w_k^{m*}[n] &:=& % w = 1_{...}
	\mathbbm{1}_{\{ (L_k^m[n]=\max_{q} L_k^q[n])~	
	\wedge~(L_k^m[n]>0)\}},~ \mathrm{with}\hspace{.3cm}\\
\label{E:opt_phi}L_k^m[n] &:=& %phi = ...
	L_{SU,k}^m[n] - \theta_k B_k^S[n],~ \mathrm{and}\hspace{.3cm} \\
\label{E:opt_ind}L_{SU,k}^m[n] &:=& %varphi = ...
	\beta^m C_k^m(h_k^m[n],p_k^{m*}[n])-\pi^mp_k^{m*}[n].\hspace{.3cm}
\end{eqnarray}
Two auxiliary variables $L_{SU,k}^m[n]$ and $L_k^m[n]$ have been defined. Such variables are useful to express the optimal \RA but also to gain insights on how the optimal \RA operates. Both variables can be viewed as \IRI which represent the reward that can be obtained if $w_k^m[n]$ is set to one. The indicator $L_{SU,k}^m[n]$ considers information only of the secondary network and represents the best achievable trade-off between the rate and power transmitted by the {\SU}. The risk of interfering the {\PU} is considered in $L_k^m[n]$, which is obtained by adding an interference-related term to $L_{SU,k}^m[n]$. Clearly, the (positive) multipliers $\pi^m$ and $\theta_k$ can be viewed as power and interference \emph{prices}, respectively. Note that \eqref{E:opt_sched} dictates that only the user with highest \IRI can access the channel. Moreover, it also establishes that if all users obtain a negative \IRI, then none of them should access the channel (in other words, an idle {\SU} with zero \IRI would be the winner during that time slot). This is likely to happen if, for example, the probability of the $k$th {\PU} being present is close to one, so that the value of $\theta_kB_k^S[n]$ in \eqref{E:opt_phi} is high, rendering $L_k^m[n]$ negative \emph{for all} $m$.

The expressions in \eqref{E:opt_pow}-\eqref{E:opt_ind} also reveal the favorable structure of the optimal \RA. The only parameters linking users, channels and instants are the multipliers $\pi^m$ and $\theta_k$. Once they are known, the optimal \RA can be found separately. Specifically: i) the power for a given user-channel pair, which is the one that maximizes the corresponding \IRI (setting the derivative of \eqref{E:opt_ind} to zero yields \eqref{E:opt_pow}), is found separately from the power for other users and channels; and ii) the optimal scheduling for a given channel, which is the one that maximizes the \IRI within the corresponding channel, is found separately from that in other channels. Since once the multipliers are known, the {\IRI}s depend only on information at time $n$, the two previous properties imply that the optimal \RA can be found separately for each time instant $n$. Additional insights on the optimal \RA schemes will be given in Sec. \ref{ss:analyzing_future:analyzing}.

Several methods to set the value of the dual variables $\pi^m$ and $\theta_k$ are available. Since, after relaxation, the problem has zero duality gap, there exists a \emph{constant} (stationary) optimal value for each multiplier, denoted as $\pi^{m*}$ and $\theta_k^*$, such that substituting $\pi^m = \pi^{m*}$ and $\theta_k = \theta_{k}^*$ into \eqref{E:opt_pow} and \eqref{E:opt_phi} yields the optimal solution to the \RA problem. Optimal Lagrange multipliers are rarely available in closed form and they have to be found through numerical search, for example by using a dual subgradient method aimed to maximize the dual function associated with \eqref{E:optimization_problem} \cite{BertsekasNP}. A different approach is to rely on stochastic approximation tools. Under this approach, the dual variables are rendered time variant, i.e., $\pi_m=\pi_m[n]$ and $\theta_k=\theta_k[n]$. The objective now is not necessarily trying to find the exact value of $\pi^{m*}$ and $\theta_k^*$, but online estimates of them that remain inside a neighborhood of the optimal value. See Sec. \ref{ss:analyzing_future:stationary_and_stochastic} and \cite{xin_convex,JSAC} for further discussion on this issue.

\subsection{RA as input for the design of the optimal sensing}\label{ss:RA:input_for_sensing}

The optimal solution in \eqref{E:opt_pow}-\eqref{E:opt_ind} will serve as input for the algorithms that design the optimal sensing scheme. For this reason, we introduce some auxiliary notation that will simplify the mathematical derivations in the next section. On top of being useful for the design of the optimal sensing, the results in this section will help us to gain insights and intuition on the properties of the optimal \RA. Specifically, let $L[n]$ be an auxiliary variable referred to as global \IRI, which is defined as
\begin{equation}\label{E:J}
L[n] := \sum\nolimits_{k} L_k[n], \;\mathrm{with} \;L_k[n]:= \sum\nolimits_{m} w_k^{m\ast}[n] L_k^m[n]
\end{equation}
Due to the structure of the optimal \RA, the \IRI for channel $k$ can be rewritten as [cf. \eqref{E:opt_sched}, \eqref{E:opt_phi}]:
\begin{equation} \label{E:J_max}
L_k[n] := \big[ \max_q {L_k^{q}[n]}\big]_{+}
\end{equation}
Mathematically, $L[n]$ represents the contribution to the Lagrangian of \eqref{E:optimization_problem} at instant $n$ when $p_k^m[n]= p_k^{m*}[n]$ and $w_k^m[n] = w_k^{m*}[n]$ for all $k$ and $m$. Intuitively, one can view $L[n]$ as the \emph{instantaneous} functional that the optimal \RA maximizes at instant $n$.

Key for the design of the optimal sensing is to understand the effect of the belief on the performance of the secondary network, thus, on $L[n]$. For such a purpose, we first define the \IRI for the {\SU}s in channel $k$ as $L_{SU,k}[n]:=\max_q L_{SU,k}^q[n]$. Then, we use $L_{SU,k}[n]$ to define the nominal \IRI vector $\boldsymbol{l}_k[n]$ as
\begin{equation}\label{E:F}
\boldsymbol{l}_k[n] :=  \binom{L_{SU,k}[n]}{L_{SU,k}[n] - \theta_k[n]}.
\end{equation}
Such a vector can be used to write $L_k[n]$ as a function of the belief $\mathbf{b}^S_k[n]$. Specifically,
\begin{equation} \label{E:J_compacto}
L_k[n] = \left[ \boldsymbol{l}^T_k [n]  \mathbf{b}^S_k[n] \right]_+.
\end{equation}
This suggests that the optimization of the sensing (which affects the value of $\mathbf{b}^S_k[n]$) can be performed separately for each of the channels. Moreover, \eqref{E:J_compacto} also reveals that $L_k[n]$ can be viewed as the expected \IRI: the second entry of $\boldsymbol{l}_k[n]$ is the \IRI if the {\PU} is present, the first entry of $\boldsymbol{l}_k[n]$ is the \IRI if it is not, and the entries of $\mathbf{b}^S_k[n]$ account for the corresponding probabilities, so that the expectation is carried over the \SIPN uncertainties. Equally important, while the value of $\mathbf{b}^S_k[n]$ is only available after making the sensing decision, the value of $\boldsymbol{l}^T_k [n]$ is available before making such a decision. In other words, sensing decisions do not have an impact on $\boldsymbol{l}^T_k [n]$, but only on $\mathbf{b}^S_k[n]$. These properties will be exploited in the next section.

%% SECCION BASIC CONCEPTS \DP

%% Secci—n \DP sensado
\section{Optimal sensing}\label{s:sensing}

The aim of this section is leveraging the results of Secs. \ref{s:problem_statement} and \ref{s:RA} to design the optimal sensing scheme. Recall that current sensing decisions have an impact not only on the current reward (cost) of the system, but also on future rewards. This in turn implies that future sensing decisions are affected by the current decision, so that the sensing decisions across time form a string of events that has to be optimized jointly. Consequently, the optimization problem has to be posed as a \DP. The section is organized as follows. First, we present a brief summary of the relevant concepts related to \DP and \POMDP which will be important to address the design of the optimal sensing for the system setup considered in this paper (Sec. \ref{ss:sensing:ReviewDP}). Readers familiar with \DP and \POMDP can skip that section. Then, we substitute the optimal \RA policy obtained in Sec. \ref{s:RA} into the original optimization problem presented in Sec. \ref{s:problem_statement} and show that the design of the optimal sensing amounts to solving a set of separate unconstrained \DP problems (Sec. \ref{ss:sensing:sensing as a DP}). Lastly, we obtain the solution to each of the \DP problems formulated (Sec. \ref{ss:sensing:solution for sensing}). It turns out that the optimal sensing leverages: $\xi_k$, the sensing cost at time $n$; the expected channel \IRI at time $n$, which basically depends on $\boldsymbol{l}_k[n]$ (\SISN) and the pre-decision belief (\SIPN); and the future reward for time slots $n'>n$. The future reward is quantified by the value function associated with each channel's \DP, which plays a fundamental role in the design of our sensing policies. Intuitively, a channel is sensed if there is uncertainty on the actual channel occupancy (\SIPN) and the potential reward for the secondary network is high enough (\SISN). The expression for the optimal sensing provided at the end of this section will corroborate this intuition.

%The last, part of the section discusses how to estimate such a value function (Sec. V.D) as well as other interesting properties of the optimal sensing scheme (Sec. V.E).

\subsection{Basic concepts about DP} \label{ss:sensing:ReviewDP}

DP is a set of techniques and strategies used to optimize the operation of discrete-time complex systems, where decisions have to be made sequentially and there is a dependency among decisions in different time instants. These systems are modeled as state-space models composed of: a set of state variables $u[n]\in\mathcal{U}$; a set of actions which are available to the controller and which can depend on the state $\alpha[n]\in \mathcal{A}(u[n])$; a transition function that describes the dynamics of the system as a function of the current state and the action taken $u[n+1] = U^{\prime} (u[n], \alpha[n], \omega[n+1])$, where $\omega[n+1]$ is a random (innovation) variable; and a function that defines the reward associated with a state transition or a state-action pair $R(u[n],\alpha[n])$. In general, finding the optimal solution of a \DP is computationally demanding. Unless the structure of the specific problem can be exploited, complexity grows exponentially with the size of the state space, the size of the action space, and the length of the temporal horizon. This is commonly referred to as the triple curse of dimensionality \cite{powell_book}. Two classical strategies to mitigate such a problem are: i) framing the problem into a specific, previously studied model and ii) find approximate solutions that allow to reduce the computational cost in exchange for a small loss of optimality.

DP problems can be classified into finite-horizon and infinite-horizon problems. For the latter class, which is the one corresponding to the problem in this paper, it is assumed that the system is going to be operated during a very large time lapse, so that actions at any time instant are chosen to maximize the expected long-term reward, i.e.,
\begin{equation}\label{E:optimalactionDP}
\max_{\alpha[n]}\E{\sum_{t=n}^\infty{\gamma^t R(u[t], \alpha[t])}}.
\end{equation}
The role of the discount factor $\gamma\in(0,1)$ is twofold: i) it encourages solutions which are focused on early rewards; and ii) it contributes to stabilize the numerical calculation of the optimal policies. In particular, the presence of $\gamma$ guarantees the existence of a stationary policy, i.e. a policy where the action at a given instant is a function of the system state and not the time instant. Note that multiplying \eqref{E:optimalactionDP} by factor $(1-\gamma)$, so that the objective resembles the one used through paper, does not change the optimal policy.

Key to solve a \DP problem is defining the so-called \emph{value function} that associates a real number with a state and a time instant. This number represents the expected sum reward that can be obtained, provided that we operate the system optimally from the current time instant until the operating horizon. If a minimization formulation is chosen, the value function is also known as \emph{cost-to-go function} \cite{BertsekasDP}. The relationship between the optimal action at time $n$ and the value function at time $n$, denoted as $V_n(\cdot)$, is given by Bellman's equations \cite{BertsekasDP,powell_book}:
\begin{subequations}\label{E:standard_Belman}
\begin{align}
V_n(u[n]) = &\max_\alpha \left\{ R(u[n],\alpha) + \mathbbm{E}_\omega \left[ V_{n+1}\left(U^{\prime}(u[n], \alpha, \omega)\right)\right]\right\}
\\
\alpha^*[n]=\alpha^*(u[n]) = \arg &\max_\alpha \left\{ R(u[n],\alpha) + \mathbbm{E}_\omega \left[ V_{n+1}\left(U^{\prime}(u[n], \alpha, \omega)\right)\right]\right\}
\end{align}
\end{subequations}
where $\omega$ is the information that arrives at time $n+1$ and thus we have to take the expectation over $\omega$. The value function for different time instants can be recursively computed by using backwards induction. Moreover, for infinite horizon formulations with $\gamma<1$, it holds that the value function is stationary. As a result, the dependence of $V_n(\cdot)$ on $n$ can  be dropped and \eqref{E:standard_Belman} can be rewritten using the stationary value function $V(\cdot)$. In this scenario, alternative techniques that exploit the fact of the value function being stationary (such as ``value iteration'' and ``policy iteration'' \cite[Ch. 2]{powell_book}) can be used to compute $V(\cdot)$.

\subsubsection{Partially Observable Markov Decision Processes}

\MDP are an important class within \DP problems. For such problems, the state transition probabilities depend only on the current state-action pair, the average reward in each step only depends on the state-action pair, and the system state is fully observable. {\MDP}s can have finite or infinite state-action spaces. {\MDP}s with finite state-action spaces can be solved exactly for finite-horizon problems. For infinite horizon problems, the solution can be approximated with arbitrary precision. A partially Observable \MDP (i.e. a \POMDP) can be viewed as a generalization of \MDP for which the state is not always known perfectly. Only an observation of the state (which may be affected by errors, missing data or ambiguity) is available. To deal with these problems, it is assumed that an \emph{observation function}, which assigns a probability to each observation depending on the current state and action, is known. When dealing with {\POMDP}s, there is no distinction between actions taken to change the state of the system under operation and \emph{actions taken to gain information}. This is important because, in general, every action has both types of effect.

The \POMDP framework provides a systematic method of using the history of the system (actions and observations) to aid in the disambiguation of the current observation. The key point is the definition of an internal \emph{belief state} accounting for previous actions and observations. The belief state is useful to infer the most probable state of the system. Formally, the belief state is a probability distribution over the states of the system. Furthermore, for {\POMDP}s this probability distribution comprises a sufficient statistic for the past history of the system. In other words, the process over belief states is Markov, and no additional data about the past would help to increase the agent's expected reward \cite{Astrom}. The optimal policy of a \POMDP agent must map the current belief state into an action. This implies that a discrete state-space \POMDP can be re-formulated (and viewed) as a continuous-space \MDP. This equivalent \MDP is defined such that the state space is the set of possible belief spaces of the \POMDP --the probability simplex of the original state space. The set of actions remains the same; and the state-transition function and the reward functions are redefined over the belief states. More details about how these functions are redefined in general cases can be found at \cite{kaelbling_POMDP}. Clearly, our problem falls into this class. The actual \SIPN is Markovian, while the errors in the \CSI render the \SIPN partially observable. These specific functions corresponding to our problem are presented in the following sections.

\subsection{Formulating the optimal sensing problem}\label{ss:sensing:sensing as a DP}

The aim of this section is to formulate the optimal decision problem as a standard (unconstrained) \DP. The main task is to substitute the optimal \RA into the original optimization problem in \eqref{E:constrained_original}. Recall that optimization in \eqref{E:constrained_original} involved variables $s_k[n]$, $w_k^m[n]$ and $p_k^m[n]$, and the sets of constraints in \eqref{E:constrained_original_shortterm_RA}, \eqref{E:constrained_original_longterm_RA} and \eqref{E:constrained_original_shortterm_Sensing}, the latter requiring $s_k[n]\in \{0,1\}$. When the optimal solution for $w_k^{m*}[n]$, $p_k^{m*}[n]$ presented in Sec. \ref{s:RA} is substituted into \eqref{E:constrained_original}, the resulting optimization problem is
\begin{subequations}\label{E:constrained_dp}
	\begin{alignat}{2} \label{E:constrained_bellman}
	\max_{\{s_k[n]\}} & ~~\bar{U}_{T|RA^*}\\
       ~\mathrm{s.~to:}~~~~& ~s_k[n]\in \{0,1\}, \label{E:constrained_bellman_binary_const}
\end{alignat}
\end{subequations}
% NOTA:
where $\bar{U}_{T|RA^*}$ stands for the total utility given the optimal \RA and is defined as
\begin{equation} \label{E:unconstrained_dp_agm}
\bar{U}_{T|RA^*} := \mathbbm{E} \left[
		 \lim_{N \to \infty} \sum_{n=0}^{N-1} (1-\gamma)\gamma^n \sum_{k}
			\Big(-\xi_k s_k[n] + \sum_{m} w_k^{m*}[n]L_k^m[n]\Big)	\right],
\end{equation}
which, using the definitions introduced in Sec. \ref{ss:RA:input_for_sensing}, can be rewritten as [cf. \eqref{E:J_max} and \eqref{E:J_compacto}]
\begin{eqnarray} \label{E:unconstrained_dp_agm_bis}
\bar{U}_{T|RA^*} &:=& \mathbbm{E} \left[
		 \lim_{N \to \infty} \sum_{n=0}^{N-1} (1-\gamma)\gamma^n \sum_{k} -\xi_k s_k[n] + L_k[n] \right] \\
&=&  \mathbbm{E} \left[
		 \lim_{N \to \infty} \sum_{n=0}^{N-1} (1-\gamma)\gamma^n \sum_{k} -\xi_k s_k[n] + \Big[
				 \boldsymbol{l}^T_k [n]  \mathbf{b}^S_k[n]
			\Big]_+ \right].
\end{eqnarray}
The three main differences between \eqref{E:constrained_dp} and the original formulation in \eqref{E:constrained_original} are that now: i) the only optimization variables are $s_k[n]$; ii) because the optimal \RA fulfills the constraints in \eqref{E:constrained_original_shortterm_RA} and \eqref{E:constrained_original_longterm_RA}, the only constraints that need to be enforced are \eqref{E:constrained_original_shortterm_Sensing}, which simply require $s_k[n]\in \{0,1\}$ [cf. \eqref{E:constrained_bellman_binary_const}]; and iii) as a result of the Lagrangian relaxation of the DP, the objective has been augmented with the terms accounting for the dualized constraints.

Key to find the solution of \eqref{E:constrained_dp} will be the facts that: i) $a_k[n]$ is independent of $h_k^m[n]$, and ii) that $a_k[n]$ is independent of $a_{k'}[n]$ for $k\neq k'$. The former implies that the state transition functions for $a_k[n]$ do not depend on $h_k^m[n]$, while the latter allows to solve for each of the channels separately. Therefore, we will be able to obtain the optimal sensing policy by solving separate {\DP}s ({\POMDP}s), which will rely only on state information of the corresponding channel. Specifically, the optimal sensing can be found as the solution of the following \DP:
\begin{equation} \label{E:unconstrained_DP_score}
	\max_{\{
	s_k[n]\in\{0,1\}\}
	}\sum_{k} \mathbbm{E} \left[
		\lim_{N \to \infty} \sum_{n=0}^{N-1}(1-\gamma)\gamma^n  \left(
			-\xi_k s_k[n] +  \Big[
				 \boldsymbol{l}^T_k [n]  \mathbf{b}^S_k[n]
			\Big]_+
		\right)
	\right],
\end{equation}
which can be separated channel-wise. Clearly, the reward function for the $k$th \DP is
\begin{equation}\label{E:reward_pomdp}
R_k [n] =  -\xi_k s_k[n] + \Big[
				\boldsymbol{l}^T_k [n]  \mathbf{b}^S_k[n]
			\Big]_+.
\end{equation}
The structure of \eqref{E:reward_pomdp} manifests clearly that this is a joint design because $s_k[n]$ affects the two terms in \eqref{E:reward_pomdp}. The first term (which accounts for the cost of the sensing scheme) is just the product of constant $\xi_k$ and the sensing variable $s_k[n]$. The second term (which accounts for the reward of the \RA) is the dot product of vectors $\boldsymbol{l}_k [n]$ (which does not depend on $s_k [n]$) and $\mathbf{b}^S_k[n]$ (which does depend on $s_k[n]$). The expression in \eqref{E:reward_pomdp} also reveals that $\boldsymbol{l}_k [n]$ encapsulates all the information pertaining the {\SU}s which is relevant to find $s_k^*[n]$. In other words, in lieu of knowing $h_k^m[n]$, $w_k^{m*}[n]$ and $p_k^{m*}[n]$, it suffices to know $\boldsymbol{l}_k[n]$.

Relying on \eqref{E:unconstrained_DP_score} and \eqref{E:reward_pomdp}, and taking into account that the problem can be separated across channels, \emph{at each time slot} $n$ the optimal sensing for channel $k$ can be obtained as [cf. \eqref{E:optimalactionDP}]
\begin{equation} \label{E:compare}
s_k^*[n]= \arg \max_{s \in \{0,1\}}	\left\{	\lim_{N \to \infty} \sum_{t=n}^{N-1}(1-\gamma )\gamma^t \mathbbm{E} \Big[ R_k[t]|s_k[n]=s \Big]  \right\}.
\end{equation}

\subsection{Bellman's equations and optimal solution}\label{ss:sensing:solution for sensing}

To find $s_k^*[n]$, we will derive the Belmman's equations associated with \eqref{E:compare}. For such a purpose, we split the objective in \eqref{E:compare} into the present and future rewards and drop the constant factor $(1-\gamma )\gamma^n$. Then, \eqref{E:compare} can be rewritten as
\begin{equation} \label{E:compare_2}
s_k^*[n]= \arg \max_{s \in \{0,1\}} \left\{	\mathbbm{E} \Big[ R_k[n]|s_k[n]=s\Big]	+ \gamma\lim_{N \to \infty} \sum_{t=n+1}^{N-1}\gamma^{t-n-1} \mathbbm{E} \Big[ R_k[t]|s_k[n]=s \Big] \right\}.
\end{equation}
It is clear that the expected reward at time slot $t=n$ depends on $s_k[n]$ --recall that both terms in \eqref{E:reward_pomdp} depend on $s_k[n]$. Moreover, the expected reward at time slots $t>n$ also depend on the current $s_k[n]$. The reason is that $\mathbf{b}^S_k[t]$ for $t>n$ depend on the $s_k[n]$ [cf. \eqref{E:gilbert_correct_s1}]. This is testament to the fact that our problem is indeed a {\POMDP}: current actions that improve the information about the current state have also an impact on the information about the state in future instants.

To account for that effect in the formulation, we need to introduce the value function $V_k(\cdot)$ that quantifies the expected sum reward on channel $k$ for all future instants. Recall that due to the fact of \eqref{E:compare} being and infinite horizon problem with $\gamma <1$, the value function is stationary and its existence is guaranteed [cf. Sec. \ref{ss:sensing:ReviewDP}]. Stationarity implies that the expression for $V_k(\cdot)$ does not depend on the specific time instant, but only on the state of the system. Since in our problem the state information is formed by the \SISN and the \SIPN, $V_k(\cdot)$ should be written as $V_k(\boB_k[n], \mathbf{h}_k[n])$. However, since $\mathbf{h}_k[n]$ is i.i.d. across time and independent of $s_k[n]$, the alternative value function $\bar{V}_k(\boB_k[n]):=$$\Eh[V_k(\boB_k[n], \mathbf{h}_k[n])]$, where $\Eh$ denotes that the expectation is taken over all possible values of $\mathbf{h}_k[n]$, can also be considered. The motivation for using $\bar{V}_k(\boB_k[n])$ instead of $V_k(\boB_k[n], \mathbf{h}_k[n])$ is twofold: it emphasizes the fact that the impact of the sensing decisions on the future reward is encapsulated into $\boB_k[n]$, and $\bar{V}_k(\cdot)$ is a one-dimensional function, so that the numerical methods to compute it require lower computational burden.

Based on the previous notation, the standard Bellman's equations that drive the optimal sensing decision and the value function are [cf. \eqref{E:compare_2}]
\begin{align}
\label{E:bellman_decision}
s_k^*[n]%(\mathbf{b}_k[n], \boldsymbol{l}_k[n])
= \arg \max_{s \in \{0,1\}} \left\{ \Ez \left[R_k[n] \big\vert s_k[n] = s \right] + \gamma  \Ez \left[ \bar{V}_k (\boB_k[n+1]) \big\vert s_k[n] = s \right] \right \}&
\\
%\end{equation}
%\begin{equation}
\label{E:bellman_value}
\bar{V}_k(\boB_k[n])= \Eh\Big[\max_{s\in \{0,1\}} \left\{ \Ez \left[R_k[n] \big\vert s_k[n]=s \right] + \gamma  \Ez\left[ \bar{V}_k (\boB_k[n+1]) \big\vert s_k[n]=s \right]\right \}&\Big],
\end{align}
where $\Ez$ denotes taking the expectation over the sensor outcomes. Equation \eqref{E:bellman_decision} exploits the fact of the value function being stationary, manifests the dynamic nature of our problem, and provides further intuition about how sensing decisions have to be designed. The first term in \eqref{E:bellman_decision}, $\Ez\left[R_k[n] \big\vert s_k[n] =s\right]$, is the expected short-term reward conditioned to $s_k[n]=s$, while the second term, $\Ez\big[ \bar{V}_k (\boB_k[n+1]) \big\vert s_k[n]=s \big]$, is the expected long-term sum reward to be obtained in all future time instants, conditioned to $s_k[n]=s$ and that every future decision is optimal. Equation \eqref{E:bellman_value} expresses the condition that the value function $\bar{V}_k(\cdot)$ must satisfy in order to be optimal (and stationary) and provides a way to compute it iteratively.

Since obtaining the optimal sensing decision $s_k^{\ast}[n]$ at time slot $n$ (and also evaluating the stationarity condition for the value function) boils down to evaluate the objective in \eqref{E:bellman_decision} for $s_k[n]=0$ and $s_k[n] = 1$, in the following we obtain the expressions for each of the two terms in \eqref{E:bellman_decision} for both $s_k[n]=0$ and $s_k[n]=1$. Key for this purpose will be the expressions to update the belief presented in Sec. \ref{ss:SystemSetup:CSI}. Specifically, expressions in \eqref{E:gilbert_predict}-\eqref{E:gilbert_correct_s1} describe how the future beliefs depend on the current belief, on the set of possible actions (sensing decision), and on the random variables associated with those actions (outcome of the sensing process if the channel is indeed sensed).

The expressions for the expected \emph{short-term reward} [cf. first summand in \eqref{E:bellman_decision}] are the following. If $s_k[n] = 0$, the channel is not sensed, there is no correction step, and the post-decision belief coincides with the pre-decision belief [cf. \eqref{E:gilbert_correct_s0}]. The expected short-term reward in this case is:
\begin{equation}\label{E:erkns0}
\Ezz{R_k[n] \big\vert s_k[n]Ê= 0} = \Big[ \boldsymbol{l}_k[n]^T \mathbf{b}_k[n]\Big]_+.
\end{equation}
On the other hand, if $s_k[n] = 1$, the expected short-term reward is found by averaging over the probability mass of the sensor outcome $z_k[n]$ and subtracting the cost of sensing
\begin{equation} \label{E:erkns1_pre}
\Ez[R_k[n] \big\vert s_k[n]Ê= 1] = -\xi_k + \sum_{z\in\{0,1\}}\Pr(z_k[n] = z \big\vert\mathbf{b}_k[n])\Big[ \boldsymbol{l}_k[n]^T \mathbf{b}_k^S(\mathbf{b}_k[n], z)\Big]_+,
\end{equation}
which, by substituting \eqref{E:gilbert_correct_s1} into \eqref{E:erkns1_pre}, yields
\begin{equation} \label{E:erkns1}
\Ez[R_k[n] \big\vert s_k[n]Ê= 1] = -\xi_k + \sum_{z\in\{0,1\}}\Big[ \boldsymbol{l}_k[n]^T \mathbf{D}_z \mathbf{b}_k[n]\Big]_+.
\end{equation}

Once the expressions for the expected short-term reward are known, we find the expressions for the expected \emph{long-term} sum \emph{reward} [cf. second summand in \eqref{E:bellman_decision}] for both $s_k[n] = 0$ and $s_k[n] = 1$. If $s_k[n] = 0$, then there is no correction step [cf. \eqref{E:gilbert_correct_s0}], and using \eqref{E:gilbert_predict} \textcolor{black}{
\begin{equation}\label{E:vs0}
\Ezz{ \bar{V}_k (\boB_k[n+1]) \big\vert s_k[n] =0} =
%\mathbbm{E}\big[ V_n(\mathbf{P}_k \mathbf{b}_k[n], \boldsymbol{l}_k[n] )\big].
\bar{V}_k\left(\Pb{\mathbf{P}_k \mathbf{b}_k[n]}\right). % Ver definicion del comando \Pb arriba
\end{equation}}
On the other hand, if $s_k[n] = 1$, the belief for instant $n$ is corrected according to \eqref{E:gilbert_correct_s1}, and updated for instant $n+1$ using the prediction step in \eqref{E:gilbert_predict} as:
\textcolor{black}{\begin{equation}\label{E:vs1}
\Ezz{ \bar{V}_k (\boB_k[n+1]) \big\vert s_k[n] =1} = \sum_{z\in\{0,1\}} \Pr(z\big\vert \mathbf{b}_k[n])\bar{V}_k\left(\Pb{\mathbf{P}_k\mathbf{b}^S_k(\mathbf{b}_k[n], z)}\right)
\end{equation}}
Clearly, the expressions for the expected \emph{long-term reward} in \eqref{E:vs0} and \eqref{E:vs1} account for the expected value of $\bar{V}_k$ at time $n+1$. Substituting \eqref{E:erkns0}, \eqref{E:erkns1}, \eqref{E:vs0} and \eqref{E:vs1} into \eqref{E:bellman_value} yields
%
%\begin{eqnarray} \label{E:final_value}
%\nonumber \bar{V}_k(\boB_k[n]) &=&  \mathbbm{E}\Bigg[ \max  \Bigg\{ \Big[ \boldsymbol{l}_k[n]^T \mathbf{b}_k[n]\Big]_+ + \gamma \bar{V}_k\left(\Pb{\mathbf{P}_k \mathbf{b}_k[n]}\right); \\
% &-&\xi_k +  \sum_{z\in\{0,1\}} \left(\Big[ \boldsymbol{l}_k[n]^T \mathbf{D}_z \mathbf{b}_k[n]\Big]_+ + \gamma \Pr(z_k[n] \big\vert \mathbf{b}_k[n]) \bar{V}_k \bigg(
%	\frac {\Pb{\mathbf{P}_k \mathbf{D}_z \mathbf{b}_k[n]}}{\mathbf{1}^T \mathbf{D}_z \mathbf{b}_k[n]}
%	\bigg)\right) \Bigg\} \Bigg],\hspace{1cm}
%\end{eqnarray}
\begin{eqnarray} \label{E:final_value}
\nonumber \bar{V}_k(\boB_k[n]) &=& \Eh\Bigg[ \max  \Bigg\{ \Big[ \boldsymbol{l}_k[n]^T \mathbf{b}_k[n]\Big]_+ + \gamma \bar{V}_k\left(\Pb{\mathbf{P}_k \mathbf{b}_k[n]}\right); \\
 &-&\xi_k +  \sum_{z\in\{0,1\}} \left(\Big[ \boldsymbol{l}_k[n]^T \mathbf{D}_z \mathbf{b}_k[n]\Big]_+ + \gamma \Pr(z_k[n] \big\vert \mathbf{b}_k[n]) \bar{V}_k \bigg(
	\frac {\Pb{\mathbf{P}_k \mathbf{D}_z \mathbf{b}_k[n]}}{\mathbf{1}^T \mathbf{D}_z \mathbf{b}_k[n]}
	\bigg)\right) \Bigg\} \Bigg].\hspace{1cm}
\end{eqnarray}
where for the last term we have used the expression for $\mathbf{b}^S_k(\mathbf{b}_k[n], z)$ in \eqref{E:gilbert_correct_s1}. Equation \eqref{E:final_value} is useful not only because it reveals the structure of $\bar{V}_k(\boB_k[n])$ but also because it provides a mean to compute the value function numerically (e.g., by using the value iteration algorithm \cite[Ch. 3]{powell_book}).

%The latter expression has the same structure than \eqref{E:bellman_value} and $\bar{V}_k(\cdot)$ can be viewed as another value function, which only depends on the state variable which is correlated along time (in this case, the occupancy belief $\mathbf{b}_k[n]$). As a result, the Value iteration algorithm can be used also to calculate the optimal value for $\bar{V}_k(\cdot)$.
%the Value iteration algorithm can be fed with \eqref{E:final_value} instead of \eqref{E:bellman_value}

Similarly, we can substitute the expressions \eqref{E:erkns0}-\eqref{E:vs1} into \eqref{E:bellman_decision} and get the optimal solution for our sensing problem. Specifically, the sensing decision at time $n$ is
\begin{equation}\label{E:final_decision}
\begin{split}
\Big[\boldsymbol{l}_k^T[n] \mathbf{b}_k[n]\Big]_+ &+ \gamma \bar{V}_k\left(\Pb{\mathbf{P}_k \mathbf{b}_k[n]}\right)
\overset{{s_k^*[n]}=0}{\underset{{s_k^*}[n]=1}\gtrless}  \\
%\max\big\{0, \boldsymbol{f}^T[n]\boldsymbol{b}[n]\big\} +\gamma V_k^{n+1}(\boldsymbol{P}\mathbf{b}[n])\overset{{s^*[n]}=0}{\underset{{s^*}[n]=1}\gtrless} - \xi + \\
- \xi_k + \sum_{z\in\{0,1\}} \Bigg(\Big[ \boldsymbol{l}^T_k[n] \mathbf{D}_z \mathbf{b}_k[n]\Big]_+ &+ \gamma \Pr(z_k[n] \big\vert \mathbf{b}_k[n]) \bar{V}_k \bigg( 	\frac {\Pb{\mathbf{P}_k \mathbf{D}_z \mathbf{b}_k[n]}}{\mathbf{1}^T \mathbf{D}_z \mathbf{b}_k[n]}	\bigg)\Bigg).
	\end{split}
\end{equation}
The most relevant properties of optimal sensing policy (several of them have been already pointed out) are summarized next: i) it can be found separately for each of the channels; ii) since it amounts to a decision problem, we only have to evaluate the long-term aggregate reward if $s_k[n] = 1$ (the channel is sensed at time $n$) and that if $s_k[n] = 0$ (the channel is sensed at time $n$), and make the decision which gives rise to a higher reward; iii) the reward takes into account not only the sensing cost but also the utility and \QoS for the {\SU}s (joint design); iv) the sensing at instant $n$ is found as a function of both the instantaneous and the future reward (the problem is a DP); vi) the instantaneous reward depends on both the current \SISN and the current \SIPN, while the future reward depends on the current \SIPN and not on the current \SISN; and vii) to quantify the future reward, we need to rely on the value function $\bar{V}_k(\cdot)$. The input of this function is the \SIPN. Additional insights on the optimal sensing policy will be given in Sec. \ref{ss:analyzing_future:analyzing}.

\section{Numerical results} \label{s:simulations}
Numerical experiments to corroborate the theoretical findings and gain insights on the optimal policies are implemented in this section. Since an \emph{RA scheme} similar to the one presented in this paper was analyzed in \cite{JSAC}, the focus is on analyzing the properties of the optimal \emph{sensing scheme}. The readers interested can find additional simulations as well as the Matlab codes used to run them in {\small $\mathrm{http://www.tsc.urjc.es/\sim amarques/simulations/NumSimulations\_lramjr12.html}$}.

The experiments are grouped into two test cases. In the first one, we compare the performance of our algorithms with that of other existing (suboptimal) alternatives. Moreover, we analyze the behavior of the sensing schemes and assess the impact of variation of different parameters (correlation of the {\PU}s activity, sensing cost, sensor quality, and average \SNR). In the second test case, we provide a graphical representation of the sensing functions in the form of two-dimensional decision maps. Such representation will help us to understand the behavior of the optimal schemes.

The parameters for the default test case are listed in Table \ref{T:params}. Four channels are considered, each of them with different values for the sensor quality, the sensing cost and the \QoS requirements. In most cases, the value of $\check{o}_k$ has been chosen to be larger than the value of $P^{MD}_k$ (so that the cognitive diversity can be effectively exploited), while the values of the remaining parameters have been chosen so that the test-case yields illustrative results. The secondary links follow a Rayleigh model and the frequency selectivity is such that the gains are uncorrelated across channels. The parameters not listed in the table are set to one.

\begin{table}[h] \caption {Parameters of the system under test.} \label{T:params}
\vspace{-.8cm}
\begin{center} \begin{tabular}{|c|c|c|c|c|c|c||c|c|}
\hline
 $k$ & \SNR  & $P^{FA}_k$ & $P^{MD}_k$ & $\mathbf{P}_k$           & $\xi_k$ & $\check{o}_k$ & $m$ & $\check{p}_m$\\ \hline
  1  & 5 dB & 0.09       & 0.08       & [0.95, 0.05; 0.02, 0.98] & 1.00    & 0.30          &  1  &  20.0 \\ \hline
  2  & 5 dB & 0.09       & 0.08       & [0.95, 0.05; 0.02, 0.98] & 1.80    & 0.05          &  2  &  16.0 \\ \hline
  3  & 5 dB & 0.05       & 0.03       & [0.95, 0.05; 0.02, 0.98] & 1.00    & 0.10          &  3  &  18.0 \\ \hline
  4  & 5 dB & 0.05       & 0.03       & [0.95, 0.05; 0.02, 0.98] & 1.80    & 0.10          &  4  &  10.0 \\ \hline
\end{tabular}
\end{center}
\end{table}
\vspace{-.3cm}

\noindent \textbf{Test case 1: Optimality and performance analysis.} The objective here is twofold. First, we want to numerically demonstrate that our schemes are indeed optimal. Second, we are also interested in assessing the loss of optimality incurred by suboptimal schemes with low computational burden. Specifically, the optimal sensing scheme is compared with the three suboptimal alternatives described next. A) A myopic policy, which is implemented by setting $V(B)=0 \; \forall B $. This is equivalent to the greedy sensing and \RA technique proposed in \cite{xin_convex}, since it only accounts for the reward of sensing in the current time slot and not in the subsequent time slots. B) A policy which replaces the infinite horizon value function with a horizon-1 value function. In other words, a sensing policy that makes the sensing decision at time $n$ considering the (expected) reward for instants $n$ and $n+1$. C) A rule-of-thumb sensing scheme implementing the simple (separable) decision function: $s_k[n] = \mathbbm{1}_{\{L_k[n]  \in [\xi_k, \theta_k-\xi_k]\}}\mathbbm{1}_{\{B_k[n] \in [\mathbf{b}^S_k(A_k, 0), \mathbf{b}^S_k(A_k, 1)]\}}$. In words, the channel is sensed if and only if the following two conditions are satisfied: a) the channel's \IRI is greater than the sensing cost and less than the interfering cost minus the sensing cost; and b) the uncertainty on the primary occupancy is higher than that obtained from a unique, isolated measurement.

\begin{figure}[htb]
	\centering
	\subfigure[]{%\hspace{-4mm}
		\includegraphics[scale=0.40]{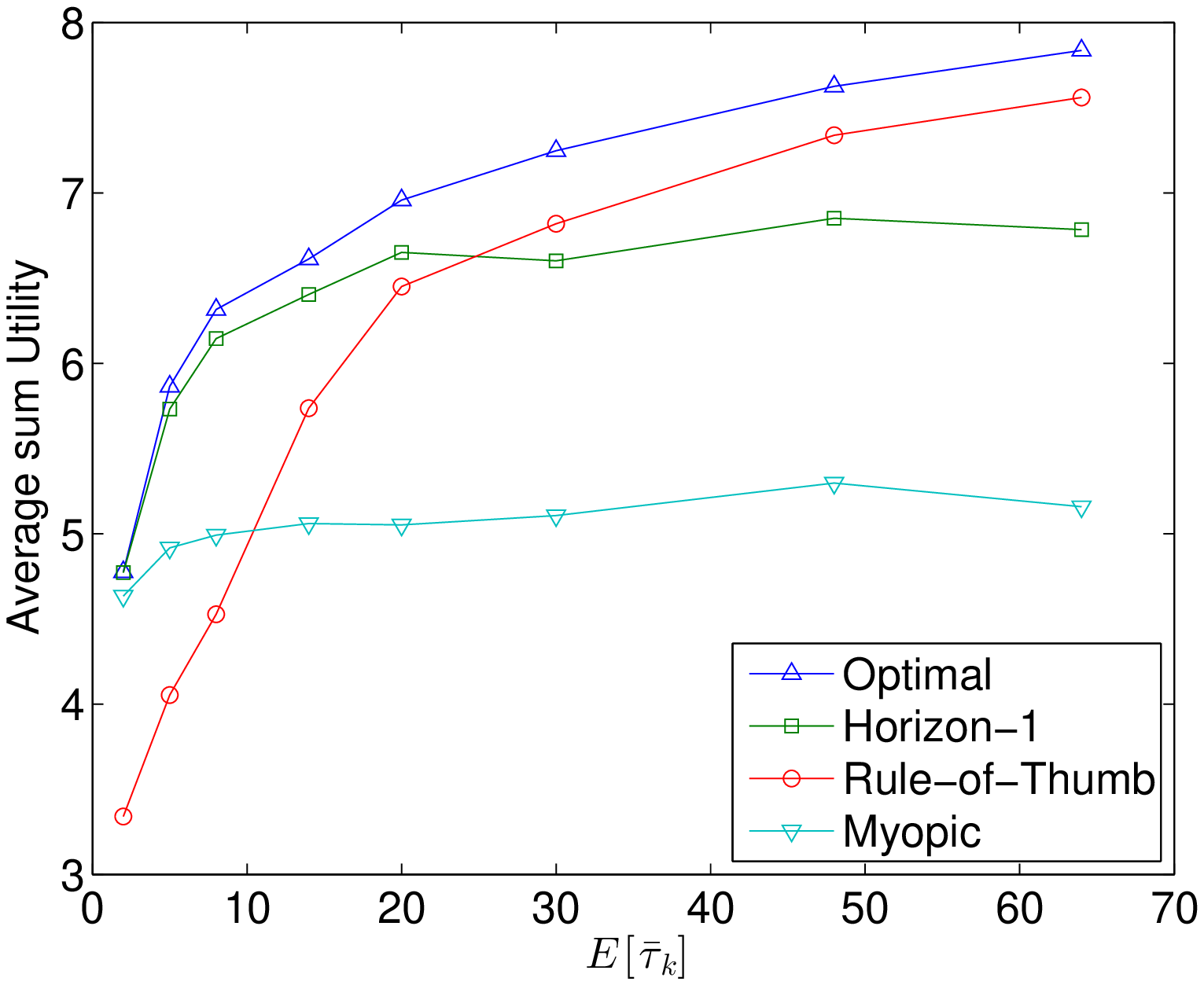}
		\label{f:1_corr}
	}
	\subfigure[]{%\hspace{-7mm}
		\includegraphics[scale=0.40]{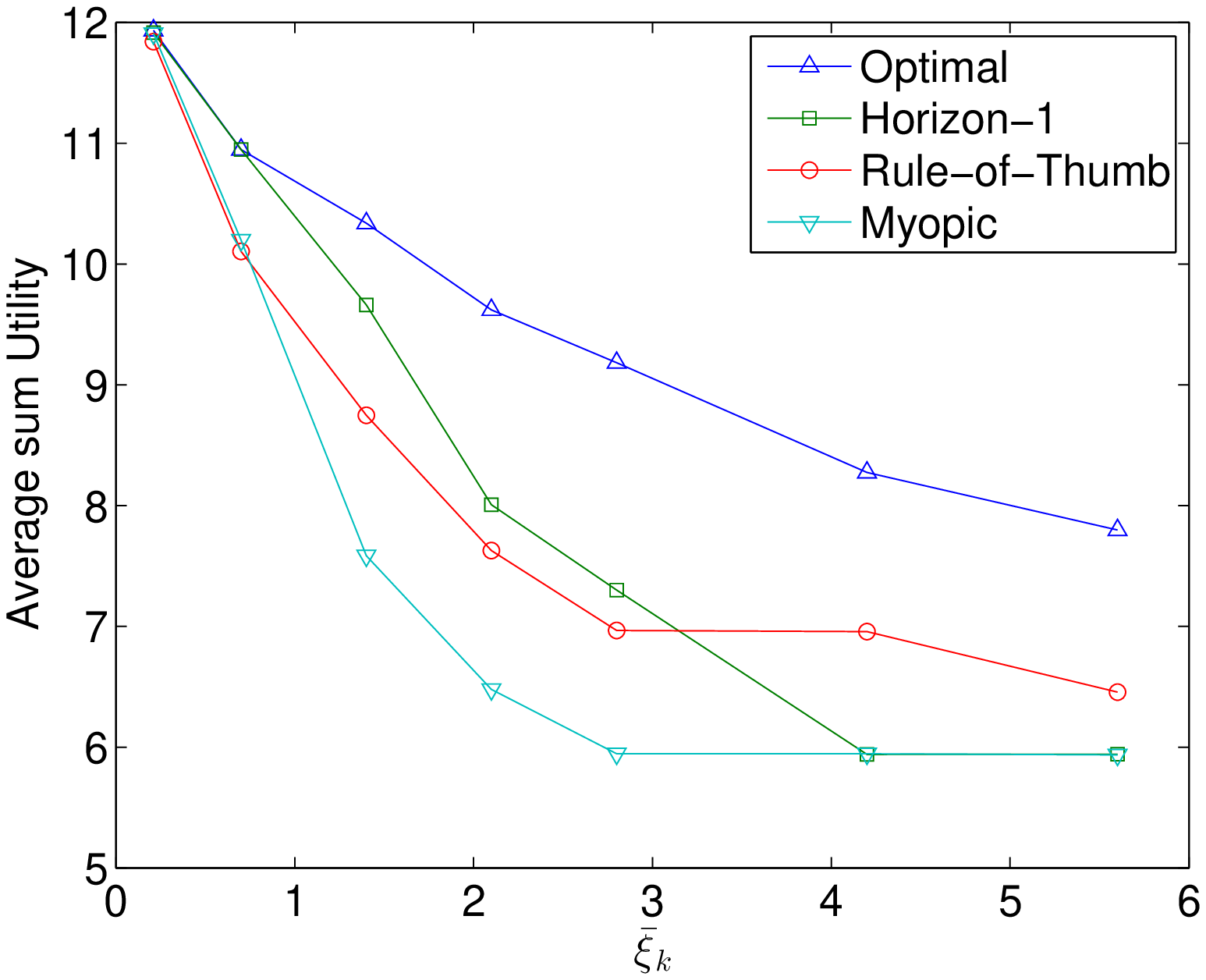}
		\label{f:2_cost}
		%\hspace{-12mm}
	}
	\\
	%\vspace{-1cm}
	\subfigure[]{%\hspace{-4mm}
		\includegraphics[scale=0.40]{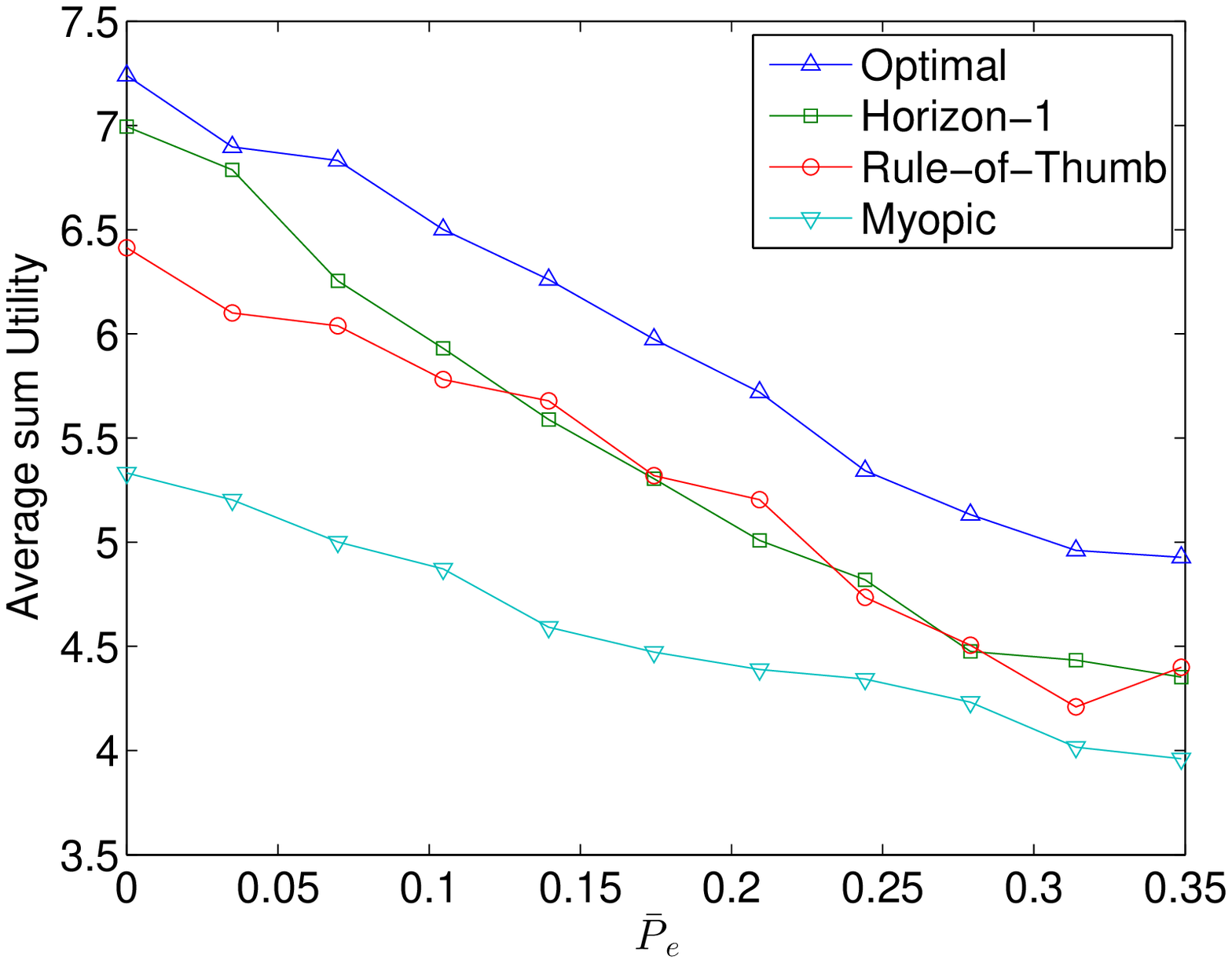}
		\label{f:3_err}
	}
	\subfigure[]{%\hspace{-7mm}
		\includegraphics[scale=0.40]{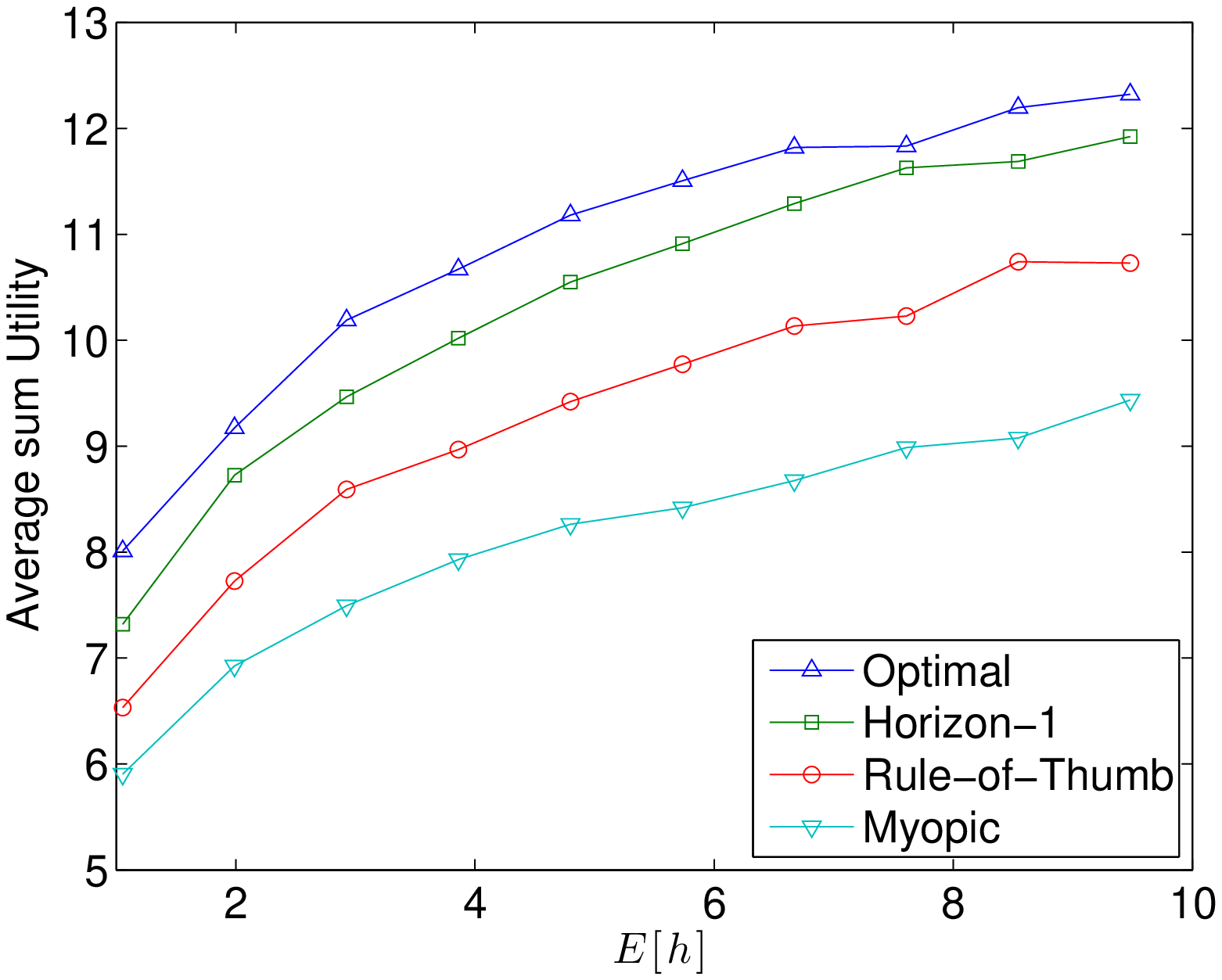}
		\label{f:4_SNR}
		%\hspace{-12mm}
	}

\vspace{-0.1cm}
\caption{{\small Performance of the optimal scheme vs. some suboptimal schemes for variations in (a) expected primary transition time; (b) sensing cost; (c) probability of error; (d) average SNR.
}}
\vspace{-0.2cm}
\label{f:comparison}
\end{figure}

Results are plotted in Fig. \ref{f:comparison}. The slight lack of monotonicity observed in the curves is due to the fact that simulations have been run using a Monte-Carlo approach. As expected, the optimal sensing scheme achieves the best performance for all test cases. Moreover, Figs. \ref{f:1_corr} and \ref{f:2_cost} reveal that the \emph{horizon-1} value function approximation constitutes a good approximation to the optimal value function in two cases: i) when the expected transition time is short (low time correlation) and ii) when the sensing cost is relatively small. The performance of the myopic policy is shown to be far from the optimal. This finding is in disagreement with the results obtained for simpler models in the opportunistic spectrum access literature \cite{zhao_framework} where it was suggested that the myopic policy could be a good approximation to solve the associated \POMDP efficiently. The reason can be that the \CR models considered were substantially different (the \RA schemes in this paper are more complex and the interference constraints are formulated differently). In fact, the only cases where the myopic policy seems to approximate the optimal performance are: i) if $\xi_k \to 0$, this is expected because then the optimal policy is to sense at every time instant; and ii) if the {\PU}s activity is not correlated across time (which was the assumption in \cite{xin_convex}).
	
Fig. \ref{f:3_err} suggests that the benefits of implementing the optimal sensing policies are stronger when sensors are inaccurate. In other words, the proposed schemes can help to soften the negative impact of deploying low quality (cheap) sensing devices. Finally, results in Fig. \ref{f:4_SNR} also suggest that changes in the average \SNR between {\SU} and \NC, have similar effects on the performance of all analyzed schemes.

\begin{figure}[h]
	%\centering
	\subfigure[]{%\hspace{-4mm}
		\includegraphics[scale=0.40]{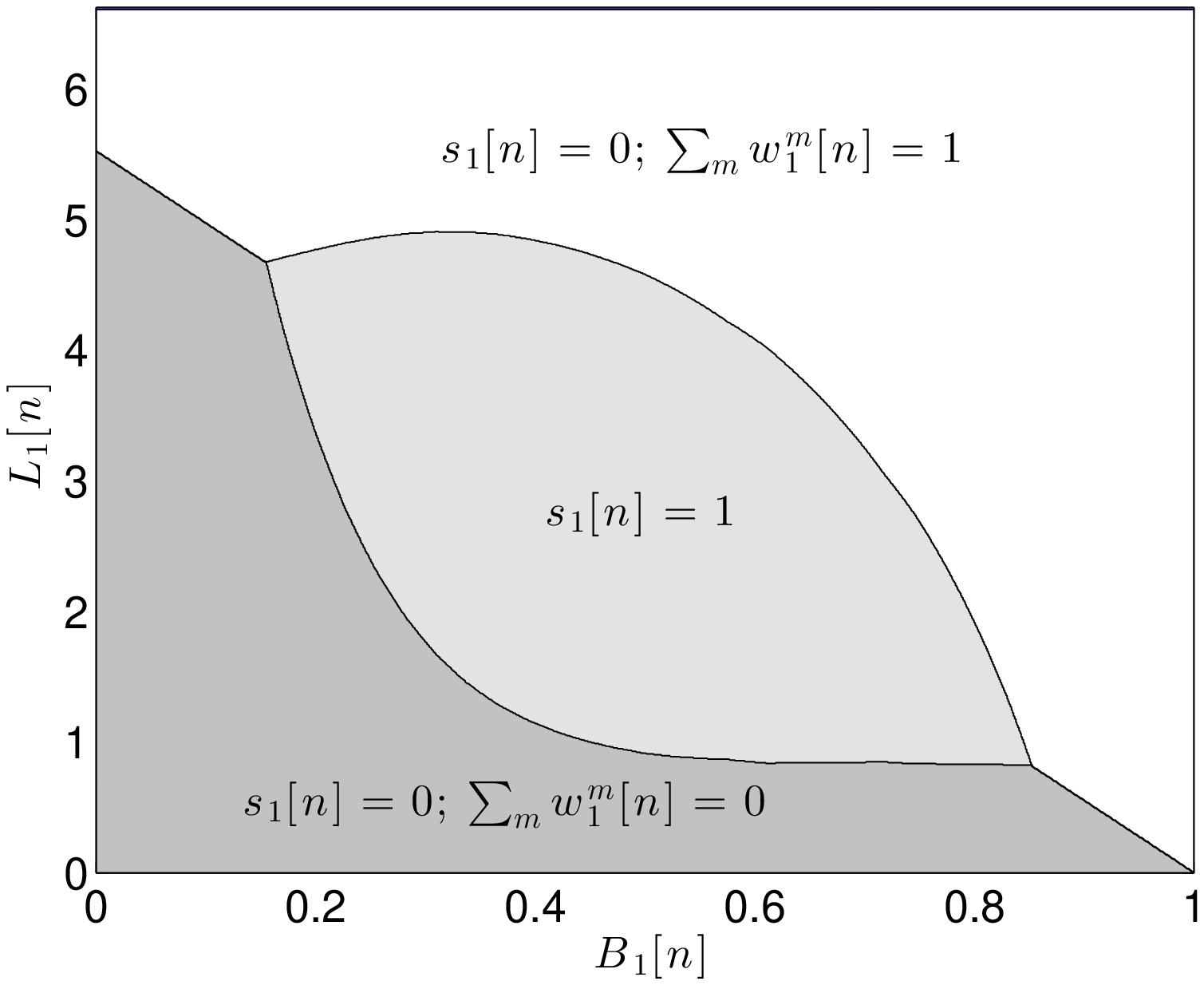}
		\label{sf:hoja11}
	}
	\subfigure[]{%\hspace{-7mm}
		\includegraphics[scale=0.40]{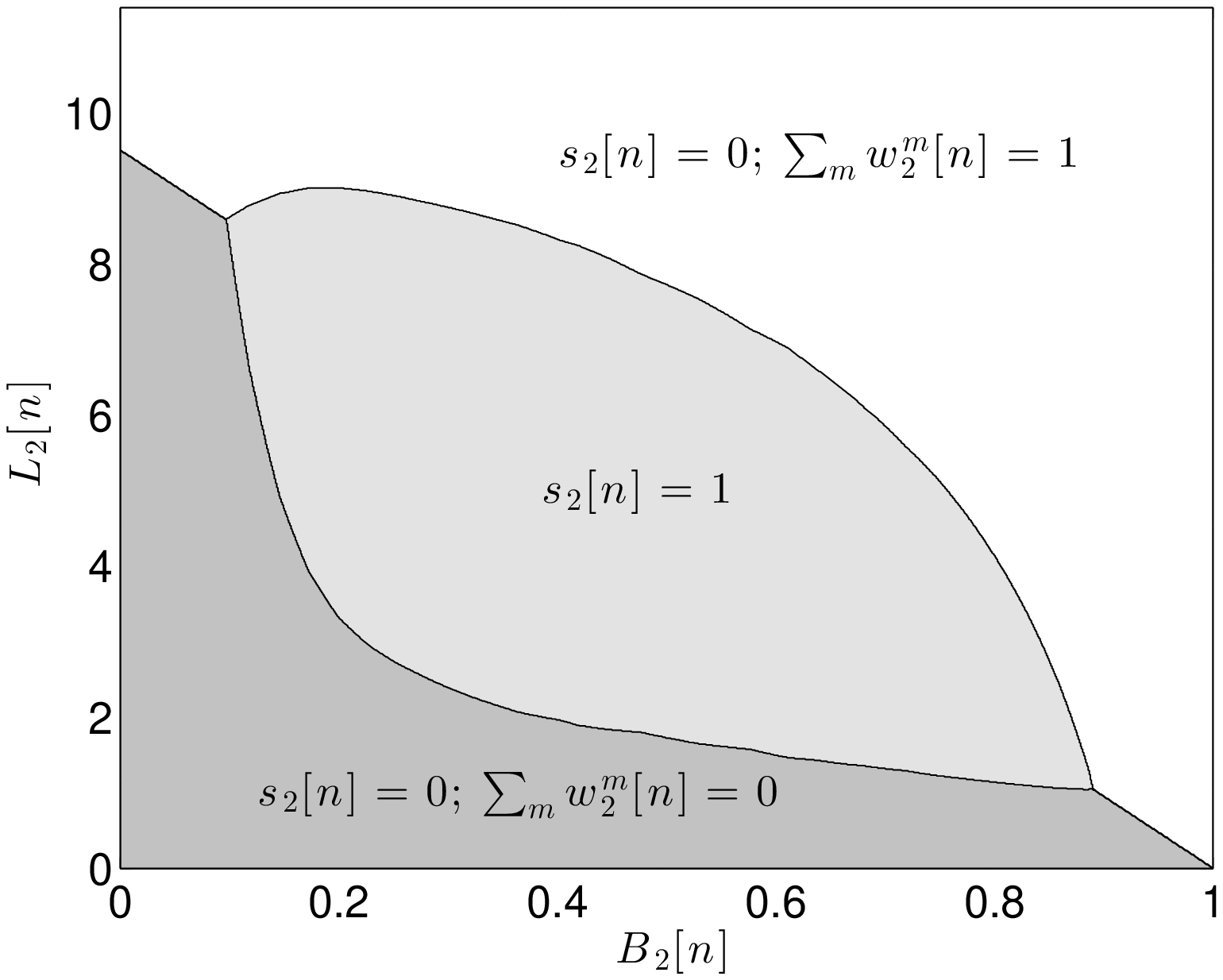}
		\label{sf:hoja12}
		%\hspace{-12mm}
	}
	\\
	%\vspace{-1cm}
	\subfigure[]{%\hspace{-4mm}
		\includegraphics[scale=0.40]{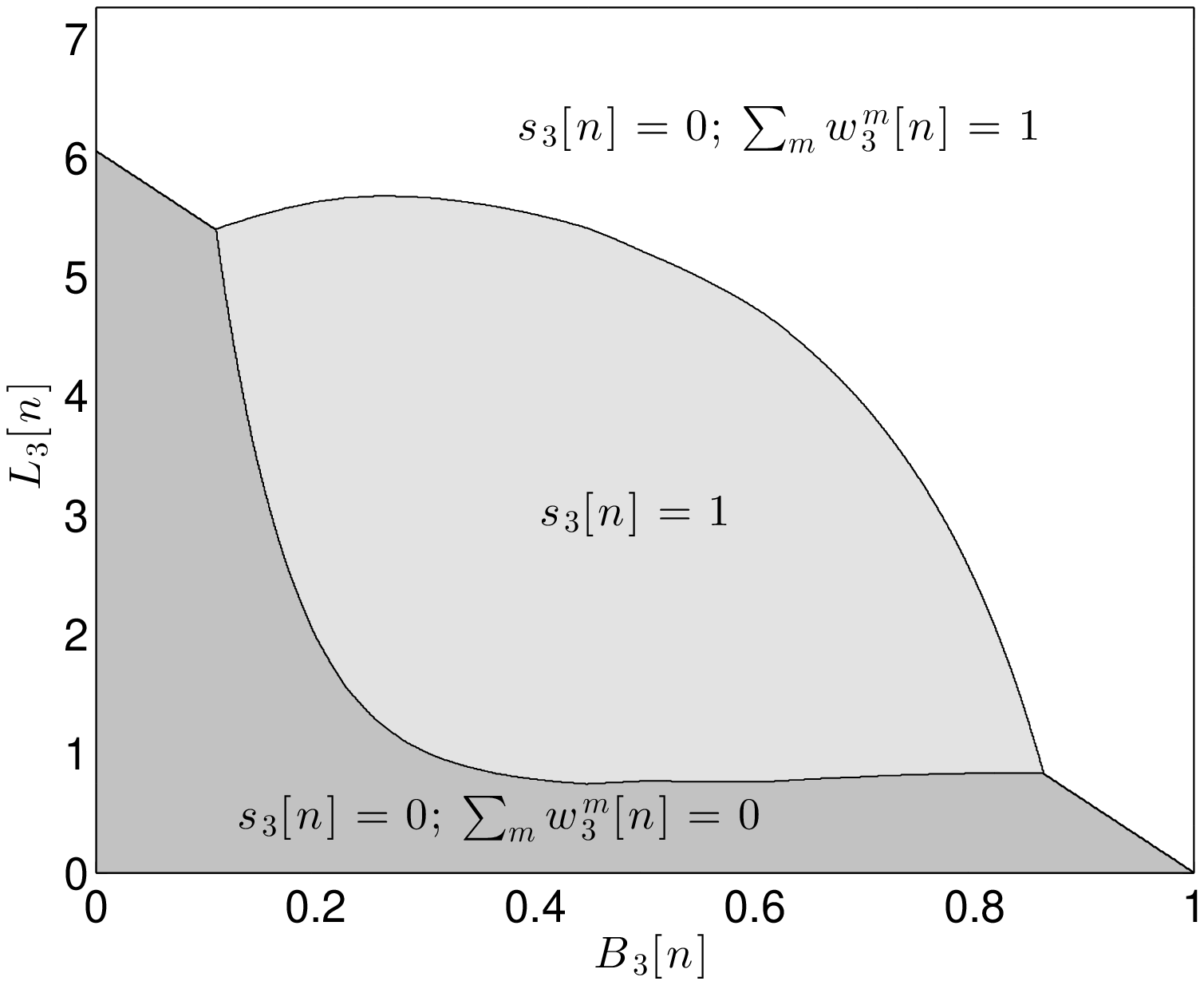}
		\label{sf:hoja21}
	}
	\subfigure[]{%\hspace{-7mm}
		\includegraphics[scale=0.40]{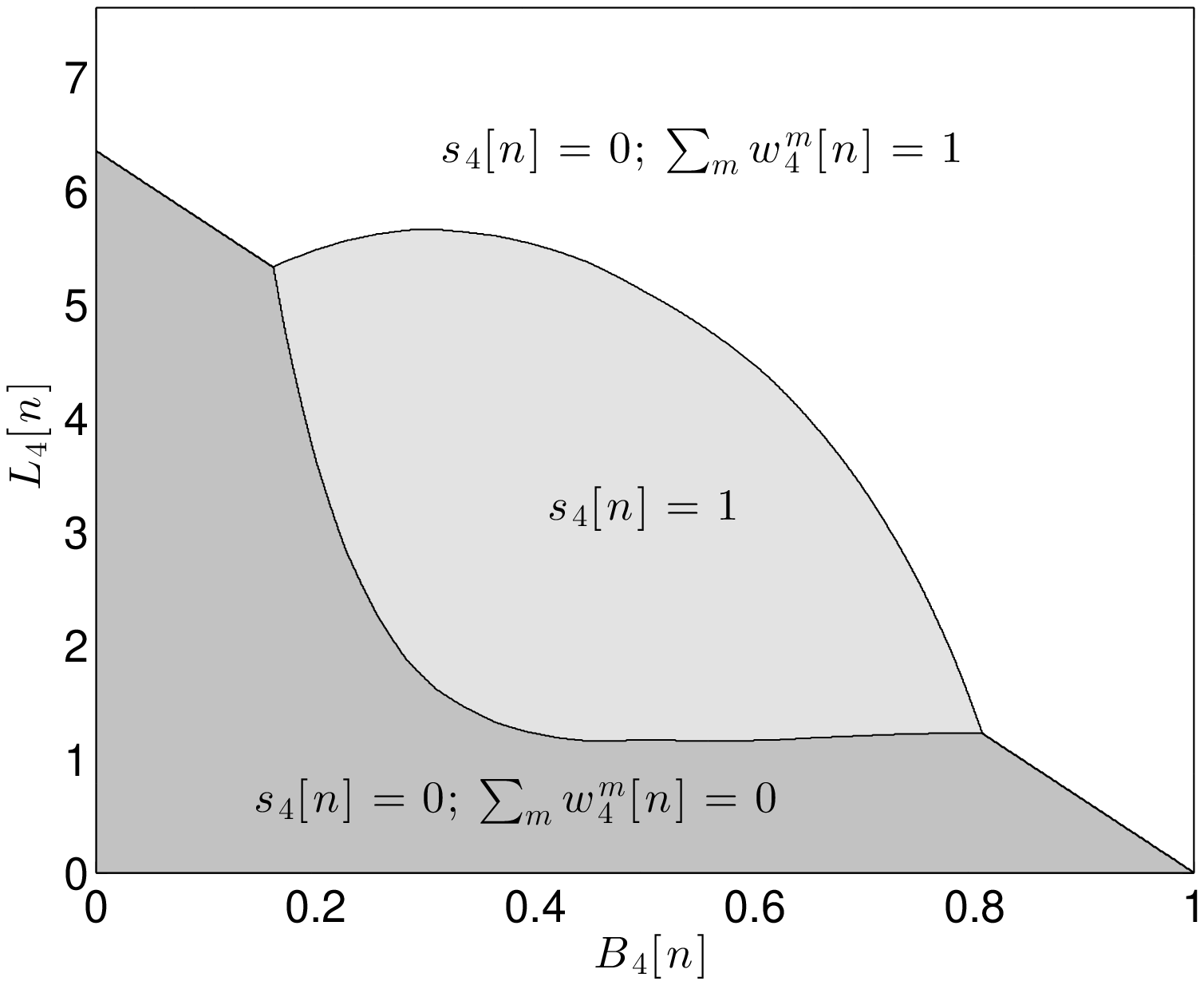}
		\label{sf:hoja22}
		%\hspace{-12mm}
	}

\vspace{-0.1cm}
\caption{{\small Decision maps (regions) for the four channels in the default test case (see Table \ref{T:params}). The light gray area in the center corresponds to the sensing decision.
}}
\vspace{-0.2cm}
\label{f:maps}
\end{figure}

\noindent \textbf{Test case 2: Sensing decision maps.} To gain insights on the behavior of the optimal sensing schemes, Fig. \ref{f:maps} plots the sensing decisions as a function of $B_k[n]$ and $L_k[n]$. Simulations are run using the parameters for the default test case (see Table \ref{T:params}) and each subplot corresponds to a different channel $k$. Since the domain of the sensing decision function is two dimensional, the function itself can be efficiently represented as an image (map). To primary regions are identified, one corresponding to the pairs $(B_k[n],L_k[n])$ which give rise to $s_k[n]=1$, and one corresponding to the pairs giving rise to $s_k[n]=0$. Moreover, the region where $s_k[n]=0$ is split into two subregions, the first one corresponding to $\sum_m w_k^{m}[n]=1$ (i.e., when there is a user accessing the channel) and the second one when $\sum_m w_k^{m}[n]=0$ (i.e., when the system decides that no user will access the channel). Note that for the region where $s_k[n]=1$, the access decision basically depends on the outcome of the sensing process $z_k[n]$ (if fact, it can be rigorously shown that $\sum_m w_k^{m}[n]=1$ if and only if $z_k[n] = 1$).

Upon comparing the different subplots, one can easily conclude that the size and shape of the $s_k[n]=1$ region depend on $\mathbf{P}_k$, $P_k^{FA}$, $P_k^{MD}$, $\xi_k$, and $\check{o}_k$.
For example, the simulations reveal that channels with stricter interference constraint need to be more frequently sensed and thus the sensing region is larger: Fig. \ref{sf:hoja11} vs. Fig. \ref{sf:hoja12}. They also reveal that if the sensing cost $\xi_k$ increases, then the sensing region becomes smaller: Fig. \ref{sf:hoja21} vs. Fig. \ref{sf:hoja22}. This was certainly expected because if sensing is more expensive, then resources have to be saved and used only when the available information is scarce.

\vspace{-1mm}

\section{Analyzing the joint schemes and future lines of work} \label{s:analyzing_future}

This section is intended to summarize the main results of the paper, analyze the properties of the optimal \RA and sensing schemes, and briefly discuss extensions and future lines of work.

\subsection{Jointly optimal RA and sensing schemes} \label{ss:analyzing_future:analyzing}

The aim of this paper was to design jointly optimal \RA and sensing schemes for an overlay cognitive radio with multiple primary and secondary users. The main challenge was the fact that sensing decisions at a given instant do not only affect the state of the system during the instant they are made, but also the state of the future instants. As a result, our problem falls into the class of \DP, which typically requires a very high computational complexity to be solved. To address this challenge efficiently, we formulated the problem as an optimization over an infinite horizon, so that the objective to be optimized and the \QoS constraints to be guaranteed were formulated as long-term time averages. The reason was twofold: i) short-term constraints are more restrictive than their long-term counterparts, so that the latter give rise to a better objective, and ii) long-term formulations are in general easier to handle, because they give rise to stationary solutions. Leveraging the long-term formulation and using dual methods to solve our \emph{constrained} sum-utility maximization, we designed optimal schemes whose input turned out to be: a) the current \SISN; b) the current \SIPN; c) the stationary Lagrange multipliers associated with the long-term constraints; and d) the stationary value (reward-to-go) function associated with the future long-term objective. While a) and b) accounted for the state of the system at the current time instant $n$, c) and d) accounted for the effect of sensing and \RA in the long-term (i.e., for instants other than $n$). In particular, the Lagrange multipliers $\pi^m$ and $\theta_k$ accounted for the long-term cost of satisfying the corresponding constraints. This cost clearly involves all time instants and cannot be computed based only on the instantaneous \RA. Similarly, the value function $\bar{V}_k(\cdot)$ quantified the \emph{future} long-term reward.

Due to a judiciously chosen formulation, our problem could be separated across channels (and partially across users), giving rise to simple and intuitive expressions for the optimal \RA and the optimal sensing policies. Specifically, for each time instant $n$, their most relevant properties were:
\begin{itemize}
\item The optimal \RA depends on: the \SISN at instant $n$, the \SIPN at instant $n$ (post-decision belief), and the Lagrange multipliers [cf. \eqref{E:opt_pow}-\eqref{E:opt_ind}]. The effect of the sensing policy on the \RA is encapsulated into the post-decision belief vector (testament to the fact that this is a joint design). The effect of other time instants is encapsulated into $\pi^m$ (long-term price of the transmission power) and $\theta_k$ (long-term price for interfering the $k$th {\PU}). \RA decisions are made so that the instantaneous \IRI is maximized. The \IRI is a trade-off between a reward (rate transmitted by the {\SU}) and a cost (compound of the power consumed by the \SU and the probability of interfering the \PU). The \RA is accomplished in a rather intuitive way: each user selects its power to optimize its own \IRI, and then in each of the channel the system picks the {\SU} who achieves the highest \IRI (so that the \IRI for that channel is maximized).
\item
The optimum sensing depends on: the \SISN at instant $n$, the \SIPN at instant $n$ (pre-decision belief), the Lagrange multipliers, and the value function [cf. \eqref{E:final_decision}]. The optimum sensing is a trade-off between the expected instantaneous \IRI (which depends on the current \SISN and \SIPN), the instantaneous sensing cost, and the future reward (which is given by the value function $\bar{V}_k(\cdot)$ and the current \SIPN). Both the instantaneous \IRI and the value function depend on the Lagrange multipliers and the \RA policies, testament to the fact that this is a joint design.
\end{itemize}
For each time instant, the \CR had to run five consecutive steps that were described in detail in Sec. \ref{s:problem_statement}. The expressions for the optimum sensing in \eqref{E:final_decision} had to be used in step T2.2, while the expressions for the optimal \RA in \eqref{E:opt_pow}-\eqref{E:opt_ind} had to be used in step T3. Once the values of $\pi^m$, $\theta_k$ and $\bar{V}_k(\cdot)$ were found (during the initialization phase of the system), all five steps entailed very low computational complexity.

\subsection{Sensing cost}\label{ss:analyzing_future:sensing_cost}

To account for the cost of sensing a given channel, the additive and constant cost $\xi_k$ was introduced. So far, we considered that the value of $\xi_k$ was pre-specified by the system. However, the value of $\xi_k$ can be tuned to represent physical properties of the \CR. Some examples follow. Example 1: Suppose that to sense channel $k$, the \NC spends a power $P_k^{NC}$. In this case, $\xi_k$ can be set to $\xi_k=\pi^{NC} P_k^{NC}$, where $\pi^{NC}$ stands for the Lagrange multiplier associated with a long-term power constraint on the NC. Example 2: Consider a setup for which the long-term rate of sensing is limited, mathematically, this can be accomplished by imposing that $\mathbbm{E} [
		 \lim_{N \to \infty} \sum_{n=0}^{N-1} (1-\gamma)\gamma^n s_k[n]]\leq \eta$, where $\eta$ represents the maximum sensing rate (say $10\%$). Let $\rho_k$ be the Lagrange multiplier associated with such a constraint, in this scenario $\xi_k$ should be set to $\xi_k=\rho_k$.  Example 3: Suppose that if the \NC senses a channel, one fraction of the slot (say 25$\%$) is lost. In this scenario $\xi_k[n]=0.25 L_k[n]$ (time-varying opportunity cost). Linear combinations and stochastic versions of any of those costs are possible too. Similarly, if a collaborative sensing scheme is assumed, aggregation of costs across users can also be considered.

\subsection{Computing the multipliers and the value functions}\label{ss:analyzing_future:stationary_and_stochastic}

In this paper, both the objective to be optimized as well as the \QoS requirements were formulated as long-term (infinite-horizon) metrics, cf. \eqref{E:constrained_original_obj}, \eqref{E:c_power} and \eqref{E:c_prob_int}. As a result, the value function associated with the objective in \eqref{E:constrained_original_obj} and the Lagrange multipliers associated with constraints \eqref{E:c_power} and \eqref{E:c_prob_int} are stationary (time invariant). Obtaining $\bar{V}_k(\cdot)$, $\pi^m$ and $\theta_k$ is much easier than obtaining their counterparts for short-term (finite-horizon) formulations. In fact, the optimum value of $\bar{V}_k(\cdot)$, $\pi^m$ and $\theta_k$ for the short-term formulations would vary with time, so that at every time instant a numerical search would have to be implemented. Differently, for the long-term formulation, the numerical search has to be implemented only once. Such a search can be performed with iterative methods which are known to converge. At each iteration those methods perform an average over the random (channel) state variables, which is typically implemented using a Montecarlo approach. Such a procedure may be challenging not only from a computational perspective, but also because there may be cases where the statistics of the random processes are not known or they are not stationary. For all those reasons, low-complexity stochastic estimations of $\bar{V}_k(\cdot)$, $\pi^m$ and $\theta_k$ are also of interest. Regarding the Lagrange multipliers, dual stochastic subgradient methods can be used as low-complexity alternative with guaranteed performance (see, e.g., \cite{xin_convex,JSAC} and references therein for examples in the field of resource allocation in wireless networks). Development of stochastic schemes to estimate $\bar{V}_k(\cdot)$ is more challenging because the problem follows into the category of functional estimation. Methods such as Q-learning \cite[Ch.8]{powell_book} or existing alternatives in the reinforcement learning literature can be considered for the problem at hand. Although design and analytical characterization of stochastic implementations of the schemes derived in this paper are of interest, they are out of the scope of the manuscript and left as future work.

\subsection{Extending the results to other CR setups}\label{ss:analyzing_future:future}

There are multiple meaningful ways to extend our results. One of them is to consider more complex models for the \CSI. Imperfect \SISN can be easily accommodated into our formulation. Non-Markovian models for the {\PU} activity can be used too. The main problem here is to rely on models that give rise to efficient ways to update the belief, e.g., by using recursive Bayesian estimation; see \cite{JSAC} and references therein for further discussion on this issue. Finally, additional sources of correlation (correlation across time for the \SISN and correlation across channels for the \SIPN) can be considered too, rendering the \POMDP more challenging to solve. Another line of work is to address the optimal design for \CR layouts different from the one in this paper. An overlay \CR was considered here, but underlay \CR networks are of interest too. In such a case, information about the channel gains between the {\SU}s and {\PU}s would be required. Similarly, in this paper we limit the interference to the {\PU} by bounding the average probability of interference. Formulations limiting the average interfering power or the average rate loss due to the interfering power are other reasonable options. Last but not least, developing distributed implementations for our novel schemes is also a relevant line of work. Distributed solutions should address the problem of cooperative sensing as well as the problem of distributed \RA. Distributed schemes should be able to cope with noise and delay in the (state) information the nodes exchange, so that a previous step which is key for developing distributed schemes is the design of stochastic versions for the sensing and \RA allocation policies. For some of this extensions, designs based on suboptimal but low complexity solutions may be a worth exploring alternative.

\bibliographystyle{IEEEtran}

\end{document}